\documentclass[11pt]{article}
\usepackage{amssymb}
\usepackage{latexsym}
\usepackage[mathscr]{eucal}
\usepackage{citesort}
\usepackage{url}
\usepackage{a4}
\usepackage{cite}
\usepackage{ntheorem}

\newcommand{\sgstatic}{{strictly globally static{}}}
\newcommand{\gstatic}{{globally static{}}}

\newcommand{\riemg}{g}
\newcommand{\riemgz}{g_0}
\newcommand{\hhat}{\ghat}
\newcommand{\DeltaL}{\Delta_{\mathrm L}}
\newcommand{\ghat}{\gamma}
\newcommand{\hzhat}{\gamma_0}

\newcommand{\del}{\partial}
\newcommand{\Mbar}{{\overline M}}

\renewcommand{\hbar}{{\overline \riemgz}}
\newcommand{\dm}{{\partial M}}

\newcommand{\letters}
  {\renewcommand{\theenumi}{\alph{enumi}}
   \renewcommand{\labelenumi}{(\theenumi)}}

\newcommand{\tp}{\tau_p}
\newcommand{\hF}{\hat F}

\newcommand{\mnu}{\nu}
\newcommand{\homega}{\hat \omega}%
\newcommand{\kerp}{\mathrm{Ker}_p\nabla X}%
\newcommand{\kerpi}{\mathrm{Inv}_p}%
\newcommand{\const}{\mathrm{const}}
\newcommand{\ml}{{M_{0,n-2\ell}}}
\newcommand{\mil}{{M_{\mathrm{iso},\ell}}}
\newcommand{\mtwo}{{M_{0,n-2}}}
\newcommand{\mfour}{{M_{0,n-4}}}

\newcommand{\selltwo}{\pi_\Sigma(\mtwo)}
\newcommand{\sfour}{\pi_\Sigma(\mfour)}
\newcommand{\stwo}{\pi_\Sigma(\mtwo)}
\newcommand{\mi}{{M_{0,n-2i}}}
\newcommand{\mj}{{M_{0,n-2j}}}
\newcommand{\htau}{{\hat \tau}}
\newcommand{\ttau}{{\tilde \tau}}
\newcommand{\zM}{{\mathring{M}}}
\newcommand{\ztau}{{\mathring{\tau}}}
\newcommand{\zSigma}{{\mathring{\Sigma}}}
\newcommand{\sings}{{\Sigma_{\mathrm{sing,iso}}}}
\newcommand{\miso}{{M_{\mathrm{iso}}}}
\newcommand{\psings}{{\partial\Sigma_{\mathrm{sing,iso}}}}
\newcommand{\sing}{{\Sigma_{\mathrm{sing}}}}
\newcommand{\singt}{{\Sigma_{\mathrm{sing},0}}}
\newcommand{\psingt}{{\partial\Sigma_{\mathrm{sing},0}}}
\newcommand{\Ein}{\operatorname{Ein}}
\newcommand{\hlambda}{\hat \lambda}

\newcommand{\lormet  }{{\frak g}}

\newcommand{\nablash}{\nabla{\kern -.75 em
     \raise 1.5 true pt\hbox{{\bf/}}}\kern +.1 em}
\newcommand{\Deltash}{\Delta{\kern -.69 em
     \raise .2 true pt\hbox{{\bf/}}}\kern +.1 em}
\newcommand{\Rslash}{R{\kern -.60 em
     \raise 1.5 true pt\hbox{{\bf/}}}\kern +.1 em}
\newcommand{\Div}{\operatorname{div}}

\newcommand{\Ric}{\operatorname{Ric}}

\newcommand{\mcO}{{\mycal O}}

\newcommand{\mcU}{{\mycal U}}

\newcommand{\hyp}{\Sigma}

\newcommand{\barg}{\bar g}

\newcommand{\threeg}{{g_\Sigma}} 

\newcommand{\mcM}{{\mycal M}}

\newcommand{\bea}{\begin{eqnarray}}
\newcommand{\beaa}{\begin{eqnarray*}}
\newcommand{\bean}{\begin{eqnarray}\nonumber}

\newcommand{\bel}[1]{\begin{equation}\label{#1}}
\newcommand{\beal}[1]{\begin{eqnarray}\label{#1}}
\newcommand{\beadl}[1]{\begin{deqarr}\label{#1}}
\newcommand{\eeadl}[1]{\arrlabel{#1}\end{deqarr}}
\newcommand{\eeal}[1]{\label{#1}\end{eqnarray}}
\newcommand{\eead}[1]{\end{deqarr}}
\newcommand{\eea}{\end{eqnarray}}
\newcommand{\eeaa}{\end{eqnarray*}}

\newcommand{\be}{\begin{equation}}
\newcommand{\ee}{\end{equation}}

\newcommand{\tr}{\mbox{\rm tr}\,}

\newcommand{\eq}[1]{\eqref{#1}}
\newcommand{\Eq}[1]{Equation~(\ref{#1})}
\newcommand{\Eqsone}[1]{Equations~(\ref{#1})}
\newcommand{\Eqs}[2]{Equations~(\ref{#1})-\eq{#2}}

\DeclareFontFamily{OT1}{rsfs}{}
\DeclareFontShape{OT1}{rsfs}{m}{n}{ <-7> rsfs5 <7-10> rsfs7 <10->
rsfs10}{} \DeclareMathAlphabet{\mycal}{OT1}{rsfs}{m}{n}
\def\scri{{\mycal I}}%
\def\Scri{\scri}

\usepackage{amsmath} 

\renewcommand{\theequation}{\thesection.\arabic{equation}}

\let\ssection=\section

\renewcommand{\section}{\setcounter{equation}{0}\ssection}

\newtheorem{defi}{\sc Definition\rm}[section]

\newtheorem{theorem}[defi]{\sc Theorem\rm}
\newtheorem{Theorem}[defi]{\sc Theorem\rm}

\newtheorem{theor}[defi]{\sc Theorem\rm}

\newtheorem{Proposition}[defi]{\sc Proposition\rm}

\newtheorem{Lemma}[defi]{\sc Lemma\!\rm}

\theorembodyfont{\upshape}
\newtheorem{Remark}[defi]{{\sc Remark}\rm}

\newtheorem{remark}[defi]{{\sc Remark}\rm}

\newcommand{\qed}{\hfill $\Box$\bigskip}

\newcommand{\proof}{\noindent {\sc Proof:\ }}

\def \Reel{\mathbb{R}}
\def \C{\mathbb{C}}
\def \R {\Reel}

\newcommand{\mcL}{{\mycal L}}
\def \Nat{\mathbb{N}}
\def \Z{\mathbb{Z}}
\def \N {\Nat}

\newcommand{\bM}{\,\overline{\!M}}

\newcounter{mnotecount}[section]

\newcommand{\rmnote}[1]{}

\begin{document}
\title{Non-trivial,  static, geodesically complete space-times with
a negative cosmological constant II. $n\ge 5$}
\author{
Michael T. Anderson\thanks{ Department of Mathematics, S.U.N.Y. at
Stony Brook, Stony Brook, N.Y. 11794-3651. Partially supported by
an NSF Grant DMS 0305865; email
\protect\url{anderson@math.sunysb.edu}}\\ Piotr T.
Chru\'{s}ciel\thanks{D\'epartement de Math\'ematiques, Facult\'e
des Sciences, Parc de Grandmont, F37200 Tours, France. Partially
supported by a Polish Research Committee grant 2 P03B 073 24;
email \protect\url{ piotr.chrusciel@lmpt.univ-tours.fr}, URL
\protect\url{ www.phys.univ-tours.fr/}$\sim$\protect\url{piotr}}\\
Erwann Delay\thanks{ D\'epartement de Math\'ematiques, Facult\'e
des Sciences, rue Louis Pasteur, F84000 Avignon, France. Partially
supported by the ACI program of the French Ministry of Research;
email
  \protect\url{erwann.delay@univ-avignon.fr}}}
\date{}

\date{}

\maketitle


\begin{abstract}
We show that the recent work of
Lee~\cite{Lee:fredholm} implies existence of a large class of new
singularity-free strictly static  Lorentzian vacuum solutions of
the Einstein equations with a negative cosmological constant. This
holds in all space-time dimensions greater than or equal to four,
and leads both to strictly static solutions and to black hole
solutions. The construction allows in principle for metrics
(whether black hole or not) with Yang-Mills-dilaton fields
interacting with gravity through a Kaluza-Klein coupling.
\end{abstract}

\section{Introduction}\label{Sintro}

In  recent work~\cite{ACD} we have constructed  a large class of
non-trivial static, geodesically complete, four-dimensional vacuum
space-times with a negative cosmological constant. The object of
this paper is to establish existence of higher dimensional
analogues of the above.

More precisely, we wish to show that  for $\Lambda<0$ and  $n\ge
4$ there exist $n$--dimensional \emph{strictly static}\footnote{We
shall say that a space-time is \emph{strictly static} if it
contains a globally timelike hypersurface orthogonal Killing
vector field.} solutions $(\mcM,\lormet )$ of the vacuum Einstein
equations with the following properties:
\begin{enumerate}
\item \label{p1}$(\mcM,\lormet   )$ is diffeomorphic to $\R\times
\hyp$, for some $(n-1)$--dimensional spacelike Cauchy surface
$\hyp$, with the $\R$ factor corresponding to the action of the
isometry group. \item\label{p2} $(\hyp,\threeg)$, where $\threeg$
is the metric induced by $\lormet   $ on $\hyp$, is a complete
Riemannian manifold. \item\label{p3} $(\mcM,\lormet   )$ is
geodesically complete. \item \label{p4}All polynomial invariants
of $\lormet $ constructed using the curvature tensor and its
derivatives up to any finite order
 are  bounded on $\mcM$.
 \item
\label{p5a} $(\mcM,\lormet   )$ admits a globally hyperbolic (in
the sense of manifolds with boundary) conformal completion with a
timelike $\Scri$. The completion is smooth if $n$ is even, and is
of differentiability class at least $C^{n-2}$ if $n$ is odd.
\item\label{p5} $(\hyp,\threeg)$ is a conformally compactifiable
manifold, with the same differentiabilities as in point \ref{p5a}.
\item \label{p6}The connected component of the group of isometries
of $(\mcM,\lormet )$ is exactly $\R$, with an associated Killing
vector $X$ being timelike throughout $\mcM$. \item \label{p7}There
exist no local solutions of the Killing equation  other than the
(globally defined) timelike Killing vector field $X$.
\end{enumerate}
An example of a manifold satisfying points \ref{p1}-\ref{p5} above
is of course $n$-dimensional anti-de Sitter solution. Clearly it
\emph{does not} satisfy points \ref{p6} and \ref{p7}.

We expect that there exist stationary \emph{and not static}
solutions as above, which can be constructed by solving an
asymptotic Dirichlet problem for the Einstein equations in a
conformally compactifiable setting. We are planning to study this
question in the future.

In a black hole context we have an obvious variation of the above;
we discuss this in more detail in Section~\ref{Sbh}.

Throughout this work we restrict attention to dimension $n\ge 4$.

Our approach is, in some sense,  opposite to that
in~\cite{Mars:1999is,Simon:1995ty}, where techniques previously
used in general relativity  have been employed to obtain
uniqueness results in a Riemannian setting.  Here we start with
Riemannian Einstein metrics and obtain Lorentzian ones by ``Wick
rotation", as follows:
Suppose that $(M,\riemg)$ is an $n$-dimensional conformally
compactifiable Einstein manifold of the form $M=\hyp\times S^1$,
and that $S^1$ acts on $M$ by rotations of the $S^1$ factor while
preserving the metric. Denote by $X=\partial_\tau$ the associated
Killing vector field, and assume that $X$ is orthogonal to the
sets $\hyp\times \{\exp(i\tau)\}$, where $\exp(i\tau)\in
S^1\subset \C$. Then the metric $\riemg$ can be (globally) written
in the form
\bel{e0}\riemg = u^2 d\tau^2 + \threeg\;,\qquad \mcL_Xu =
\mcL_X\threeg=\threeg(X,\cdot)=0\;.\ee It is straightforward to
check that the space-time $(\mcM:=\R\times\Sigma,\lormet   )$,
with
\bel{e0n}\lormet   = -u^2 dt^2 + \threeg\ee
is a static solution of the vacuum Einstein equations with
negative cosmological constant, with Killing vector field
$\partial_t$.

In order to continue, some definitions are in order: Let $M$ be
the interior of a smooth, compact, $n$-dimensional
manifold-with-boundary $\Mbar$. A Riemannian manifold $(M,g)$ will
be said to be \emph{conformally compact at infinity}, or
\emph{conformally compact}, if
 $$ g=x^{-2}\barg\;,$$
 for a smooth function $x$ on $
 \Mbar$ such that $x$ vanishes precisely on
 the boundary $\partial \Mbar$ of $\Mbar$, with non-vanishing gradient
 there. Further $\barg$ is a Riemannian metric which is regular
 up-to-boundary on $\Mbar$; the differentiability properties near $\partial M$ of a conformally compact metric $g$
 will always refer to those of $\barg$. The operator $$P:=\DeltaL+2(n-1)\;,$$ where $\DeltaL$ is the
Lichnerowicz Laplacian (\emph{cf., e.g.,}~\cite{Lee:fredholm})
associated with $\riemg$, plays an important role in the study of
such metrics. An Einstein metric $g$ with scalar curvature
$-n(n-1)$ will be said {\em non-degenerate} if $P$ has trivial
$L^2$ kernel on the space of trace-free symmetric $2$-tensors. We
prove the following openness theorem around static metrics for
which the Killing vector field has no zeros (see
Section~\ref{Stsa} for terminology):

\begin{Theorem}\label{Tmain}  Let
$(M, g)$ be a non-degenerate, strictly globally static conformally
compact Riemannian Einstein metric, with conformal infinity
$\gamma:=[\bar g|_{\partial M}]$ (conformal equivalence class).
Then any small static perturbation of $\gamma$ is the conformal
infinity of a strictly globally static Riemannian Einstein metric
on M.
\end{Theorem}

 Theorem~\ref{Tmain} is established by the arguments presented at the beginning of Section~\ref{SsoluRtoL},
 compare~\cite{ACD} for a more detailed treatment.
In Section~\ref{SsoluRtoL}  we also describe a subclass of
Riemannian metrics given by Theorem~\ref{Tmain} that leads to
Lorentzian Einstein metrics
with the properties \ref{p1}-\ref{p7} listed above. 
In particular we show that our construction provides non-trivial
solutions near the  AdS solution in all dimensions.

By an abuse of terminology, Riemannian solutions for which the set
of zeros of the Killing vector field $X$ contains an $n-2$
dimensional surface $N\equiv N^{n-2}$ will be referred to as
\emph{black hole solutions}; $N$ will be called\footnote{In the
corresponding Lorentzian solution the set $N$ will be the
pointwise equivalent of the \emph{black hole bifurcation surface},
\emph{i.e.}, the intersection of the past and future event
horizons.} \emph{the horizon}. The corresponding openness result
here reads:

\begin{Theorem}\label{Tmainbh} Let
$(M, g)$ be a non-degenerate, globally static conformally compact
Riemannian Einstein metric,  with strictly globally static
conformal infinity $\gamma$. Suppose that \begin{enumerate}
\item either $M=N\times \R^2$, with the action of $S^1$ being by rotations of $\R^2$, or
\item $H_{n-3}(M)=\{0\}$, and the zero set of $X$ is a smooth
$(n-2)$--dimensional submanifold with trivial normal bundle.
\end{enumerate} Then any small globally static
perturbation of $\gamma$ is the conformal infinity of a static
black hole solution with horizon $N$.
\end{Theorem}

The proof of Theorem~\ref{Tmainbh} is given at the end of
Section~\ref{Sbh}.

 This paper is organised as
follows: In Section~\ref{SsoluR} we review some results concerning
the Riemannian equivalent of the problem at hand. In
Section~\ref{SsoluRtoL} we sketch  the construction of the new
solutions, and we show that our results in the remaining sections
prove existence of non-trivial solutions which are near the
$n$-dimensional anti-de Sitter one. In Sections~\ref{Sbh} and
\ref{SKK} we discuss how the method here can be used to produce
solutions with black holes, and with Kaluza-Klein type coupling to
matter. In Section~\ref{Sho} we give conditions under which
hypersurface-orthogonality descends from the boundary to the
interior: this is done under the hypothesis of \emph{topological
staticity} in Section~\ref{Stsa}, and under the hypothesis of
\emph{existence of a twist potential} in Section~\ref{Swtp}.
Appendix~\ref{Aext} studies the action of Killing vector fields
near the conformal boundary, and contains results about
extendibility of conformal isometries of $\partial M$ to
isometries of $M$. In Appendix~\ref{Ante} we derive the norm and
twist equations for Einstein metrics in all dimensions; those
equations are of course well known in dimension four. In
Appendix~\ref{Asofp} we study the structure of the metric near
fixed points of the action of the isometry group, as needed for
the analysis of Section~\ref{Sho}. In Appendix~\ref{Sndbhs} we
calculate the sectional curvatures of $n$-dimensional Kottler-type
solutions, proving in particular the existence of a large family
of non-degenerate $n$-dimensional black hole solutions, for any
$n$.

\section{The solutions}\label{Ssolu}

We start by a review of the associated Riemannian problem:
\subsection{The Riemannian solutions}\label{SsoluR}

An important result on the structure of conformally compact
Einstein manifolds is the following --- an improvement of earlier
results in~\cite{GL,biquard}  (\emph{cf.} also
\cite{mand1,mand2,erwann3} for related or similar results):

\begin{theorem}[Lee~\cite{Lee:fredholm}]\label{thm:einstein} Let $M$ be the interior of a
smooth, compact, $n$-dimensional manifold-with-boundary $\Mbar$,
$n\ge 4$, and let $\riemgz$ be a non-degenerate Einstein metric on
$M$ that is conformally compact of class $C^{l,\beta}$ with $2\le
l\le n-2$ and $0<\beta<1$. Let $\rho$ be a smooth defining
function for $\del M$, and let $\hzhat = \rho^2 \riemgz|_{\del
M}$. Then there is a constant $\epsilon>0$ such that for any
$C^{l,\beta}$ Riemannian metric $\ghat$ on $\dm$ with
$\|\ghat-\hzhat\|_{C^{l,\beta}}<\epsilon$, there exists an
Einstein metric $\riemg$ on $M$ that has $[\ghat]$ as conformal
infinity and is conformally compact of class $C^{l,\beta}$.
\end{theorem}

 The non-degeneracy condition above will hold {\em e.g.}\/ in the
following circumstances:

\begin{theorem}\label{thm:einstein2}
Under the remaining hypotheses of Theorem~\ref{thm:einstein},
$\DeltaL+2(n-1)$ has trivial  $L^2$ kernel on the space of
trace-free symmetric $2$-tensors if either of the following
hypotheses is satisfied:
\begin{enumerate}\letters
\item At each point, either all the sectional curvatures of $g_0$
are nonpositive, or all are bounded below by $-2(n-1)/n$.
  \item The Yamabe invariant of $[\hhat]$ is nonnegative and $\riemgz$ has
sectional curvatures bounded above by $(n-1)(n-9)/8(n-2)$.
\end{enumerate} Here
$g_0$ has been normalised so that its scalar curvature equals
$-n(n-1)$.
\end{theorem}

Point (b) of Theorem~\ref{thm:einstein2} is due to
Lee~\cite{Lee:fredholm}. Point (a) is easily inferred from Lee's
arguments, for completeness we give the proof in
Appendix~\ref{Sndbhs}.
%
%
%
%
%

The regularity of the solutions above can be improved as
follows~\cite{CDLS}; the following result is the only exception to
the rule that $n\ge 4$ in this paper:

\begin{theorem}\label{treg1}  Let $g$ be a $C^{2}$ conformally compactifiable Einstein
metric on an $n$-dimensional manifold $M$ with $C^\infty$ smooth
boundary metric $[\gamma]$, $n\ge 3$.
\begin{enumerate}\item If $n$ is even or equal to three, then there exists a differentiable structure on
$\bM$ such that $g$ is smoothly compactifiable. \item If $n$ is
odd, then there exist local coordinates $(x,v^C)$  near the
boundary so that
\bel{A00}g=x^{-2}(dx^2 +\ghat_{AB}dv^A dv^B)
\;,\ee with the functions $\ghat_{AB}$ of the form
\bel{A0} \ghat_{AB}(x,v^C)=\phi_{AB}(x,v^C,x^{n-1}\ln x)\;,\ee
with $\phi_{AB}(x,v^C,z)$ --- smooth functions of all their
arguments. Further,  there exists a differentiable structure on
$\bM$ so that $g$ is smoothly compactifiable if and only if
$(\partial_z\phi_{AB})(0,v^C,0)$ vanishes.
\end{enumerate}
\end{theorem}

Explicit formulae for $(\partial_z\phi_{AB})(0,v^C,0)$ in low
dimensions can be found in~\cite{deHaro:2000xn}. It follows from
the results in~\cite{FG} that $(\partial_z\phi_{AB})(0,v^C,0)=0$
when $\ghat(0)$ has a representative which is Einstein, so that
the filling metric $g$ is smoothly compactifiable in this case,
independently of the dimension.

\subsection{From Riemannian to Lorentzian solutions}\label{SsoluRtoL}

 In order to implement the procedure leading from \eq{e0} to
 \eq{e0n}, we start  by solving the
Einstein equation for $\riemg$ with a prescribed conformal
infinity $[\ghat]$ on $\partial M =
\partial \hyp\times S^1$, using \emph{e.g.}\/
Theorem~\ref{thm:einstein}. One further assumes that rotations of
the $S^1$ factor are conformal isometries of $[\ghat]$. In order
to carry through the construction one needs to know  that
conformal isometries of $[\ghat]$ extend to isometries of
$\riemg$. 
%
This is proved in~\cite{mand2} in dimension four without further
restrictions, and under a non-degeneracy condition in higher
dimensions; we give an alternative
 proof of this fact in Appendix~\ref{Aext}. A somewhat weaker
version of the extension result in Appendix~\ref{Aext} has been
independently proved, using essentially the same argument, by
Wang~\cite{MR1879811}; compare~\cite{Qing:uniqueness} for yet
another independent similar result. We include the details of our
proof, because in the course thereof we derive some properties of
Killing vector fields which are used elsewhere in the paper. The
final step is to ensure hypersurface-orthogonality of the $U(1)$
action. This is done in the next section.

We refer the reader to~\cite{ACD} for a detailed analysis of the
four-dimensional case, where considerably stronger results are
available. However, even in dimension four the results on
hypersurface-orthogonality in Section~\ref{Swtp} do not follow
from those in~\cite{ACD}.

In order to show that the intersection of the set of hypotheses of
our results below is not empty,  let $(M,g_0)$ be the Riemannian
equivalent of the $n$-dimensional anti-de Sitter space-time, with
$g_0$ obtained by reversing the procedure described above. Thus,
$M$ is diffeomorphic to $B^{n-1}\times S^1$, where $B^{n-1}$ is
the $(n-1)$--dimensional open unit ball. Let $\alpha$ be any
strictly positive function on the unit $(n-2)$--dimensional sphere
$S^{n-2}$, and consider the following metric $\gamma$ on
$S^{n-2}\times S^1$: \bel{gc0} \gamma= \alpha^2
d\varphi^2+h\;,\qquad \mcL_{\partial_\phi} h =
h(\partial_\phi,\cdot)=0\;.\ee Here $\varphi$ is the coordinate
along the $S^1$ factor of $\partial M$.  Since $g_0$ has negative
sectional curvatures, by Lee's theorem~\ref{thm:einstein2} there
exists an Einstein metric on $M$ with conformal infinity
$[\gamma]$ provided that $\alpha$ is sufficiently close to one and
$h$ is sufficiently close to the unit round metric.\footnote{For
definiteness we only consider metrics close to anti-de Sitter,
though an identical argument can be used whenever the results
of~\cite{mand2}, or those of~\cite{Lee:fredholm}, apply. In
particular, in view of the results of~\cite{mand2}, in dimension
four the construction described here applies in much larger
generality.} By Proposition~\ref{killglob} $\partial_\varphi$
extends to a Killing vector field $X$ on $M$. Since the
corresponding Killing vector field in $(M,g_0)$ did not have any
zeros, continuous dependence of solutions of \eq{e1.0} upon the
metric implies that the same will hold for $X$ (making $\alpha$
closer to $1$ and $h$ closer to the unit round metric if
necessary). In fact, this also shows that the orbit space $M/S^1$
will be a smooth manifold, diffeomorphic to $B^{n-1}$. The fact
that $H_{n-3}(B^{n-1})=\{0\}$ implies existence of the twist
potential $\tau$, and since the boundary action is hypersurface
orthogonal we can use Theorem~\ref{T3} to obtain
hypersurface-orthogonality throughout $M$. Therefore the
Riemannian solutions so obtained lead to Lorentzian equivalents,
as described above.

Let us justify the claims made in the Introduction. Point~\ref{p1}
follows immediately from the discussion around \Eq{e0}.
Point~\ref{p5} follows from Theorem~\ref{treg1} and from what is
said in the proof of Theorem~\ref{T3}. Point~\ref{p2} is a
straightforward corollary of point~\ref{p5}. Point~\ref{p4}
follows from well known properties of conformally compactifiable
metrics. The geodesic completeness of the static metrics so
obtained has been proved in~\cite[Section~4]{ACD}.  Global
hyperbolicity in point~\ref{p5a} is established in the course of
the proof of Theorem~4.1 of~\cite{ChruscielSimon}, while the
differentiability properties claimed in point~\ref{p5a} follow
from point~\ref{p5}. The argument given at the end
of~\cite[Section~4]{ACD} gives non-existence of other global or
local Killing vector fields when the boundary metric has no other
conformal isometries.

\subsection{Black hole solutions}\label{Sbh}

Let us pass to a discussion of static black hole solutions in
higher dimensions.
     The standard examples of static Riemannian AdS-type black hole solutions are on the
manifold $M = N^{n-2} \times \R^2$, with $N:=N^{n-2}$ compact, and
with metric of form:
   \bel{bhform}   g_m = V^{-1} dr^2 + V d\theta^2 + r^2 g_N\;,\ee
where $g_N$ is any Einstein metric, $Ric_{g_{N}} = \lambda g_N$,
with $g_N$ scaled so that $\lambda = \pm (n-3)$ or $0$. Then for
$V = V(r)$ given by
       \bel{bhpot} V = c + r^2 - (2m)/r^{n-3}\;,\ee
with $c =\pm 1$ or $0$ respectively, $g_m$ is static Einstein,
with $Ric_{g_m} = -(n-1)g_m$. These are just the analogues of
Kottler metrics in higher dimensions. The length of $S^1\ni
\theta$ is  determined by $m$ together with the requirement that
$g_m$ be a smooth metric at the ``horizon" $r = r_0$, the largest
root of $V(r)$. (This restriction of course disappears in the
Lorentzian setting).

    Now each such $g_m$ belongs to a 1-parameter family of metrics, parameterised
by the mass $m$. One expects that for generic values of $m$, the
metric $g_m$ is non-degenerate, as defined in the introduction. In
fact, when $n = 4$, the AdS-Schwarzschild metric has non-trivial
kernel exactly for one specific value of $m$,\footnote{In
dimension four it is known  that all continuous isometries descend
from the boundary to the interior~\cite{mand2}. When infinity is
spherical, the elements of the kernel have to be spherically
symmetric, and one can conclude using the (generalised) Birkhoff
theorem.} while the toroidal black holes, as well as the higher
genus Kottler black holes, are always non-degenerate. Those last
two results are well known, and in any case are proved in
Appendix~\ref{Sndbhs}, where we also show existence of a large
class of non-degenerate black hole
 solutions for all $n\ge 5$.

    Assuming non-degeneracy, suppose we then consider local
perturbations of the conformal infinity, preserving the static
structure at infinity --- so, \emph{e.g.}, just vary  the
function, say $\alpha$, which describes the length of the $S^1$'s
at infinity, keeping the remainder of the conformal boundary
metric fixed. Then by the results in Appendix~\ref{Aext}, we get
extension of the isometric $S^1$ action on the (locally unique)
Einstein filling metric of Theorem~\ref{thm:einstein}. We need to
prove the extension is static also, this will follow if we can use
Theorems~\ref{t 1.2} or \ref{T4}. For the former, we can use the
fact that \emph{topological staticity}, as defined at the
beginning of Section~\ref{Stopsa}, is stable under continuous
deformations of the metric (compare~\cite[Lemma~2.6]{ACD}; the
restriction $\dim M=4$ there is not needed). This proves
Theorem~\ref{Tmainbh} under the hypothesis that the action of
$S^1$ is by rotations of $\R^2$, as that action is topologically
static.

Another condition which can lead to staticity is $H_{n-3}(M) = 0$,
for then we have existence of the twist potential.
    After a small change of $\gamma$, the $S^1$ action is again a small perturbation of
the original static $S^1$ action. This means that the only fixed
point set of the $S^1$ action (the zero set of $X$), is  again a
smooth  $(n-2)$ manifold, say  $N$, with the normal $S^1$ bundle
remaining trivial. Hence, condition~\eq{me8.0a} of
Theorem~\ref{T4} is satisfied, and staticity follows.

\subsection{Kaluza-Klein solutions}\label{SKK}

Similarly one should be able to construct space-times as described
in the introduction, with or without black hole regions, which
satisfy the Einstein-Yang-Mills-dilaton field equations with a
Kaluza-Klein coupling~\cite{Coquereaux:1988ne}; solutions
belonging to this family have been numerically constructed
in~\cite{Bjoraker:2000qd,Winstanley:2001bs}. More precisely,
suppose that one has a conformally compactifiable Einstein
manifold $(M^{n},g)$ satisfying  the following: a) $\partial
M^{n}= S^1\times M^{n-2}$; with a product boundary conformal class
$[\tilde g|_{\partial M^{n}}]$ such that b) rotations along $S^1$
are conformal isometries; c) $(M^{n},g)$ satisfies the hypotheses
of Theorem~\ref{thm:einstein} (compare
Theorem~\ref{thm:einstein2}); d) the $S^1$ action on $M$
associated with the rotations of $S^1$ on $\partial M$ satisfies
the hypotheses of one of the Theorems~\ref{t 1.2}, \ref{T3},
\ref{T4} or \ref{T4n}. Then any connected Lie group $G$ of
conformal isometries of $[\tilde g|_{ M^{n-2}}]$ with a free
action will then lead to a Kaluza-Klein type Yang-Mills gauge
group for the associated Lorentzian solutions.

\section{Hypersurface-orthogonality}\label{Sho}
We use the notations of Appendix~\ref{Aext}. We wish to show that
$X$ is hypersurface-orthogonal; clearly a necessary condition for
that is that $\hat X(0)$ be hypersurface orthogonal. We suspect
that this condition is sufficient, but we have not been able to
prove that. In dimension four hypersurface-orthogonality has been
proved in~\cite{ACD} under the hypothesis of \emph{topological
staticity} of $X$, as defined there, {\em cf.}\/ below.  We shall
show in Section~\ref{Stopsa} that the result generalises to higher
dimensions. We also give an alternative approach, assuming that we
have the  \emph{twist potential} at our disposal.

In what follows we will assume that $\hat X(0)$ arises from an
$S^1$ action on $\partial M$; this hypothesis can be replaced by
the existence of a hypersurface $\mycal S$  in
$M\setminus\{g(X,X)=0\}$ which is transversal to $X$
--- identical proofs apply, with $\Sigma$ there
replaced by $\mycal S$.

Before proceeding further we have to introduce some notation.
 Let $\Sigma $ be the orbit space of the $S^{1}$ action, and let $\zSigma$ denote the set of orbits of principal
type, then $\zSigma$ is a smooth manifold forming an open dense
subset of $\Sigma$. We set
$$\zM:=(\pi^{*}_{\Sigma})^{-1} \zSigma\;.$$
Let $g_{\Sigma}$ be the induced metric on $\zSigma ;$ thus the
metric $g$ has the form
\begin{equation} \label{e2}
g = u^{2}(d\phi  + \theta )^{2} + \pi^{*}_{\Sigma}g_{\Sigma},
\end{equation}
where $\theta $ is a connection 1-form, $u$ is the length of the
Killing field $X = \partial /\partial\phi $, and
$\pi^{*}_{\Sigma}:M\to\Sigma$ is the canonical projection. The
parameter $\phi $ parameterises a circle $S^{1}.$ The space
$\Sigma $ is in general a  $(n-1)$-orbifold; there may be
stratified submanifolds in $\Sigma $ along which the metric has
cone singularities.
  Of course $\Sigma $ is non-compact
--- it has a boundary at infinity $\partial_{\infty}\Sigma $
corresponding to the orbit space of the $S^{1}$ action on
$\partial M$. Redefining the $S^1$ action if necessary, one can
without loss of generality assume that the action is free on
$(\pi^{*}_{\Sigma})^{-1}\zSigma$. The set of non-principal orbits
is the union of trivial orbits $\singt$ which correspond to fixed
points of the action, and of special orbits $\sings$ which are
circles with non-trivial isotropy group:
$$\sing:=\singt\cup\sings\;,\qquad M_0:=(\pi^{*}_{\Sigma})^{-1}\singt=\{u=0\}\subset M\;.$$
In dimension four the fixed point set consists of isolated points
and smooth, totally geodesic submanifolds,  those have been called
``nuts" and ``bolts";
 nut fixed points are isolated points in $M$, while the bolts correspond to totally geodesic
surfaces in $\partial\zSigma$, \emph{cf.}\/~\cite{Gibbons:1979xm}.
The structure of the orbits near fixed points is discussed in all
dimensions in Appendix~\ref{Asofp}, see
also~\cite{Fintushel1,Fintushel2}. Off the fixed point set
the isotropy group is finite and so the orbits are circles. The
isotropy group may change. For example, in dimension three one can
have $S^{1}$ actions on a solid torus $D^{2}\times S^{1}$ which
are free on $(D^{2}\setminus
\{0\})\times S^{1},$ 
with non-trivial isotropy on the core curve $\{0\}\times S^{1}.$
One can take such and product with $S^{1}$ again to obtain higher
dimensional manifolds which have (isolated) curves where the
isotropy jumps up~\cite{0246.57017,Fintushel1,Fintushel2}.

Let the \emph{twist $(n-3)$--form} $\hat \omega$ be defined on $M$
by the equation (the definition here differs by a factor of 2 from
the definition in~\cite{ACD})
$$\hat\omega = *_g( \xi\wedge d\xi)\;, \qquad \xi:=
g(X,\cdot)\;,$$ where $*_g$ is the Hodge duality operator with
respect to the metric $g$. On $\zM$ the form $\hat\omega$ is the
pull-back by $\pi_\Sigma$ of a form $\omega$ defined on $\zSigma$.
It is well known in dimension four, and it is shown in general in
Appendix~\ref{Ante}, that $\omega$ (and hence $\hat \omega$) is
closed.

\subsection{Topologically static actions}\label{Stsa}\label{Stopsa}

  We will use the following terminology, as in~\cite{ACD}.
   The $S^{1}$ action on $(M, g)$ is {\it \sgstatic}\ if
$(M, g)$ is globally a warped product of the form
\begin{equation} \label{e1.3}
M = S^{1}\times \Sigma\;, \quad g = u^{2}d\phi^{2} +
\pi_\Sigma^{*}g_{\Sigma}\;,
\end{equation}
where $u: \Sigma \rightarrow {\Bbb R}$ is strictly positive and
$g_{\Sigma}$ is a complete metric on $\Sigma$, $\partial \Sigma =
\emptyset$. In this case, the $S^{1}$ action is just given by
rotations in the $S^{1}$ factor. The $S^{1}$ action is {\it
\gstatic} if \eq{e1.3} holds with $u = 0$ somewhere. In this case,
the locus $\{u = 0\}$ is not empty, but there are no exceptional
orbits. Next, the $S^{1}$ action is {\it topologically static} if
the $S^{1}$ bundle $S^{1} \rightarrow P \rightarrow \Sigma_{P}$ is
a trivial bundle, i.e. it admits a section. (Here $P$ is the union
of principal orbits, while we use the symbol $E$ for the union of
exceptional ones.) This is equivalent to the existence of a
cross-section of the $S^{1}$ fibration $P \cup E \rightarrow
\Sigma_{P\cup E}$. \footnote{$S$ will be called a cross-section of
a fibration if $S$ meets every fiber at least once, with the
intersection being transverse.} Finally, we  define the $S^{1}$
action to be {\it locally static} if every point of $(M, g)$ has a
neighborhood isometric to a neighborhood of a point with metric of
the form \eq{e1.3}; this is equivalent to the usual notion of
static {in the sense of} the existence of a hypersurface
orthogonal Killing field.

 We shall use an obvious equivalent of the above for an $\R$ action
by isometries on a Lorentzian manifold $(\mcM,\lormet)$, with the
further restriction that the associated Killing vector field be
timelike almost everywhere.

 The main result of this section is the following:
\begin{theorem} \label{t 1.2}
  Let $(M, g)$ be a smoothly compactifiable Einstein metric on $M$, with {\rm dim}\,$M\ge 4$.
Suppose the free $S^{1}$ action at conformal infinity $(\partial
M, \gamma )$ is \sgstatic, i.e.
\begin{equation} \label{e1.4}
\partial M = S^{1}\times V,
\end{equation}
and the $S^{1}$ action on $(M, g)$ is topologically static.

  Then the $S^{1}$ action on $(M, g)$ is \emph{locally static}, i.e.
$(M, g)$ is locally of the form \eq{e1.3}, (with $\{u = 0\} \neq
\emptyset$ possibly).
\end{theorem}

\proof The method of  proof is identical to that in~\cite{ACD}. As
pointed out in~\cite{Qing:uniqueness}, regardless of dimension and
signature one has the identity\footnote{The identity here differs
by a factor of 2 from the one in~\cite{ACD} because the form
$\omega$ here is twice that in~\cite{ACD}.}
\bel{neweq} d\left(\frac{\xi \wedge \hat\omega}{u^2}\right) = \pm \frac
{|\hat \omega|^2 i_X( \mbox{\rm Vol})}{u^4}\;,\ee with the sign
$\pm$ being determined by the signature of the metric. Here Vol is
the volume form. Integrating \eq{neweq} over any cross-section
$\Sigma$ one will obtain $\hat \omega=0$ provided that the
boundary term arising from the left-hand-side of \eq{neweq}
vanishes. This requires sufficiently fast fall-off of $\hat
\omega$ near the conformal infinity $\partial M$, which is
provided by the following Lemma. The hypotheses of Theorem~\ref{t
1.2} imply that the Killing vector has no zeros on $\partial M$,
but we do not need this assumption for the proof that
follows:

\begin{Lemma} \label{Lkil} In the coordinate system of \eq{kill0} we
have
\beaa & \hat \omega_{rA_1\ldots A_{n-4}}= O(r)
\;,\qquad \hat \omega_{AA_1\ldots A_{n-4}}= O(1)\;, &\eeaa with
$\hat \omega_{rA_1\ldots A_{n-4}}=\omega_{rA_1\ldots A_{n-4}}$ and
$\hat \omega_{AA_1\ldots A_{n-4}}=\omega_{AA_1\ldots A_{n-4}}$.
\end{Lemma}

\proof
The idea of the proof is to use the equation $$
\nabla^k\nabla_kX_i=-\Ric(g)_i{}^jX_j\;,$$ together with the fact
that $g$ is Einstein, to obtain more information about the decay
of the relevant components of the metric. We work in the
coordinate system of Appendix~\ref{Aext}, and use the conventions
there. We have (recall that
$\hat{X}_r\equiv 0$ by Remark \ref{Re2})
 \bean 
2\nabla^k\nabla_kX_A&=&\nabla^k(\nabla_k X_A-\nabla_AX_k)\\
\nonumber&=&\nabla^k(\partial_k X_A-\partial_AX_k)\\
\nonumber
&=&r^2\Big\{\partial^2_r{X}_A-\Gamma^r_{rr}\partial_r{X}_A-\Gamma^E_{rA}\partial_r{X}_E
+\hat{g}^{EF}\Big[\partial_E\partial_F{X}_A-\partial_E\partial_A{X}_F\\
\nonumber &&\hspace{2cm}-\Gamma^r_{EF}\partial_r{X}_A
-\Gamma^C_{EF}(\partial_C{X}_A-\partial_A{X}_C)\\
 &&\hspace{2cm}+\Gamma^r_{EA}\partial_r{X}_F
-\Gamma^C_{EA}(\partial_F{X}_C-\partial_C{X}_F)\Big]  \Big\}
\;.\eeal{st3} Consider any point $p$ on the conformal boundary at
which $\hat X(0)$ does not vanish. As $\hat{X}(0)$ is hypersurface
orthogonal, we can choose a local coordinate system on the
boundary at infinity, defined on a neighborhood of $p$, such that
$x^A=(\varphi,x^a)$ ($a=3,...,n$), with
$\hat{X}(0)=\partial_{\varphi}$ and
$$
\hat{g}(0)_{AB}dx^Adx^B=\hat{g}(0)_{\varphi\varphi}d\varphi^2+\hat{g}(0)_{ab}dx^adx^b.
$$
Recall that
$X_A=r^{-2}\hat{g}_{AB}\hat{X}^B=r^{-2}\hat{g}_{A\varphi}.$ From
\eq{Chr1}--\eq{Chr2} and \eq{st3} we then have
$$\begin{array}{ll}
2\nabla^k\nabla_kX_A&=2(n-1)r^{-2}\hat{g}_{A\varphi}-n
r^{-1}\hat{g}'_{A\varphi}+r^{-1}\hat{g}^{CD}(2\hat{g}'_{DA}\hat{g}_{C\varphi}
-\hat{g}'_{CD}\hat{g}_{A\varphi})\\
&\hspace{1cm}+\hat{g}''_{A\varphi}+\frac{1}{2}\hat{g}^{CD}(-2\hat{g}'_{DA}
\hat{g}'_{C\varphi}+\hat{g}'_{CD}\hat{g}'_{A\varphi})
+2\hat{\nabla}^E\hat{\nabla}_E\hat{X}_A\;,
\end{array}
$$
where a prime denotes an $r$-derivative. {}From equation
\eq{killmix3} we have
$$\hat{g}'_{A\varphi}=\hat{g}^{CD}\hat{g}'_{DA}\hat{g}_{C\varphi}$$
and then as $\Ric(g)=-(n-1)g$,
$$\begin{array}{l}
(2-n)r^{-1}\hat{g}'_{A\varphi}+\hat{g}''_{A\varphi}\\
\hspace{2cm}=r^{-1}\hat{g}^{CD} \hat{g}'_{CD}\hat{g}_{A\varphi}
-\frac{1}{2}\hat{g}^{CD}(-2{\hat{g}'_{DA}\hat{g}'_{C\varphi}}+\hat{g}'_{CD}\hat{g}'_{A\varphi})
-2\hat{\nabla}^E\hat{\nabla}_E\hat{X}_A.
\end{array}
$$
Consider that equation when $A=a$; straightforward but tedious
algebra shows that  its right-hand-side  can be written as a
linear combination of the $\hat{g}_{b\varphi}$'s together with
their first derivatives and their second
$\partial_c$--derivatives, with bounded, differentiable
up-to-boundary, coefficients built out of the $g_{AB}$'s and their
derivatives. (For example, \beaa
\hat{g}^{CD}{\hat{g}'_{Da}\hat{g}'_{C\varphi}}&=&\hat{g}^{\varphi\varphi}{\hat{g}'_{\varphi
a}\hat{g}'_{\varphi\varphi}} +\hat{g}^{c\varphi}{\hat{g}'_{\varphi
a}\hat{g}'_{c\varphi}} +\hat{g}^{c\varphi}{\hat{g}'_{c
a}\hat{g}'_{\varphi\varphi}}+
\hat{g}^{cd}{\hat{g}'_{da}\hat{g}'_{d\varphi}}\;,\eeaa and note
that $\hat g^{c\varphi}$ is a rational function involving the
$g_{b\varphi}$'s which vanishes when the latter do, hence can be
written as an expression linear in the $g_{b\varphi}$'s with
coefficients which depend upon the $g_{AB}$'s.) Then as
$\hat{g}_{a\varphi}(0)=0$ we have that
$\hat{g}_{a\varphi}=O(r^2)$. Taylor expanding and matching
coefficients in front of powers of $r$ one is led to
\bel{DLgtrans}\hat{g}_{a\varphi}=O(r^{n-1}).\ee

Set
$$\mu^2:=\hat g(0)(\hat X(0),\hat X(0))\;.$$
Since $g$ is Einstein we have~\cite{FG}  $\hat g(r) = \hat
g(0)+O(r^2)$, so that \bel{se0} u^2:= g(X,X)= r^{-2} \hat g (\hat
X,\hat X) = r^{-2}(\mu^2+O(r^2))\;.\ee In particular $u$ behaves
as $1/r$ (recall that we are so far working away from the zero set
of $\hat X(0)$). Consider the two-form $\hlambda$ defined in
\eq{nke0.6}; \eq{DLgtrans} gives \bel{se00} \hlambda =
\sum_aO(r^{n-5})dr\wedge dx^a + \sum_{a,b}O(r^{n-4})dx^a\wedge
dx^b\;.\ee The coordinates $(r,x^a)$ can be used as local
coordinates on the quotient manifold $\Sigma$, in those
coordinates $\hlambda_{ab}=\lambda_{ab}$,
$\hlambda_{ar}=\lambda_{ar}$, $\hlambda_{i\varphi}=0$. It follows
now that in any coordinate system $\{x^A\}$ on the conformal
boundary as in \eq{kill0}, not necessarily adapted to the
hypersurface-orthogonal character of $X$, we will have
\bel{se1} \lambda_{AB}=O(r^{n-4})\;,\qquad
\lambda_{Ar}=O(r^{n-5})\;,\ee as long as we are not at a point at
which $\hat X(0)$ vanishes. Now, $u\hlambda$ is defined and smooth
regardless of zeros of $X$, which implies that \eq{se1} holds
globally on each domain of definition of the coordinates $x^A$,
independently of the existence  of zeros of $\hat X(0)$ there. We
finally obtain
\bea\lambda_{AB}&=& O(r^{n-4})\quad \Longleftrightarrow \quad
\lambda^{AB}= g^{AC}g^{BD}\lambda_{CD}=O(r^{n})\;, \eeal{nke0b}
with a similar equivalence for $\lambda_{Ar}$. Let
$\eta_{A_1\ldots A_{n-2}}$ be totally anti-symmetric, equal to $1$
if $A_1\ldots A_{n-2}$ is an even permutation of $1,2,\ldots,
n-2$. It is clear from \eq{kill2}-\eq{e0.0} that
$$\sqrt{\det g_{\Sigma}}=r^{-(n-1)}\sqrt{\det \hat
g_\Sigma(r)}\;,$$ where $\hat g_\Sigma(r)$ is the metric on the
quotient $(\{r=\const\},\hat g(r))/S^1$. Choosing a convenient
orientation, from the definition \eq{nke8} of $\omega$ we have
\bean\omega_{rA_1\ldots A_{n-4}}&=&\sqrt{\det
g_{\Sigma}}\,\eta_{A_1\ldots A_{n-4}BC}\lambda^{BC}
\\&=&r^{-(n-1)}\sqrt{\det \hat g_\Sigma(r)}\,\eta_{A_1\ldots
A_{n-4}BC}\lambda^{BC}\nonumber
\\ & =& O(r)\;,\eeal{nke0c}
as desired. The claim about $\omega_{AA_1\ldots A_{n-4}}$ is
established by a similar calculation.

\hfill\qed

 Returning to
the proof of Theorem~\ref{t 1.2}, suppose first that the Killing
vector field $X$ has no zeros, and that all orbits are of
principal type. Let $S$ be a hypersurface transverse to $X$, let
$S(r)$ denote the intersection of $S$ with the level sets of the
function $r$ of \eq{kill0}, we then have
\bel{erwint}\int_S\frac{1}{u^4}|\hat \omega|^2*\xi=\pm\lim_{r\rightarrow
0}\int_{S(r)}\frac{1}{u^2}\xi\wedge\hat \omega\;,\ee with the
$\pm1$ factor as in \eq{neweq}. In local coordinates of \eq{kill0}
we have $\xi=g(X,.)=O(r^{-2})$ and $u^2\geq c r^{-2}$, while $\hat
\omega=O(1)$ by Lemma~\ref{Lkil}. Further, if we choose $S$ to be
asymptotically orthogonal to $X$, then  the pull-back of $\xi$ to
$S$ will be $o(r^{-2})$. Thus we obtain
$$\int_S\frac{1}{u^4}|\hat \omega|^2*\zeta=\lim_{r\rightarrow 0} o(1)=0.$$
Now $*\xi=\alpha|K|dvol_S$, where $\alpha$, the angle between $S$
and $K$, is of constant sign. We can conclude $\hat \omega=0$.

When zeros of $X$ occur, \eq{erwint} becomes
\bel{erwint2}\int_S\frac{1}{u^4}|\hat \omega|^2*\xi=\pm\lim_{r\rightarrow
0}\int_{S(r)}\frac{1}{u^2}\xi\wedge\hat \omega
\mp\lim_{\epsilon\rightarrow
0}\int_{\{\rho=\epsilon\}}\frac{1}{u^2}\xi\wedge\hat \omega\;.\ee
Here $\rho=\sqrt{\sum_{i=1}^\ell \rho_i^2}$, with $\rho_i$ as in
\eq{afp3}. To finish the proof, we need to show that the boundary
integral corresponding to the zeros of $X$ vanishes. We use the
coordinate system of \eq{afp0}: we have $c^{-1} \rho \le u \le c
\rho$ by \eq{afp4a.1}. Then $\hat \omega = O(\rho)$ by \eq{afp29}
and $\xi = O(\rho)$ by \eq{afp7}, hence the integrand in the
right-hand-side of \eq{erwint2} is uniformly bounded as
$\epsilon\to 0$. By scaling (or by a direct calculation, using the
formulae of Appendix~\ref{Asofp}) one sees then that the integral
vanishes at least as fast as $\epsilon^{2\ell-1}$, whence the
result. \qed

 The above result is reasonably satisfactory from a general
relativistic point of view: in that case the solutions of main
interest possess spacelike hypersurfaces transverse to the Killing
vector field, which imply topological staticity of the associated
Riemannian solution. Nevertheless, it seems of interest to look
for other hypotheses which will lead to hypersurface-orthogonality
of the Killing vector. In the next section we will obtain some
such results under the hypothesis that there exists a \emph{twist
potential} $\tau$, \emph{i.e.},\/ $ \omega=d\tau$.

\subsection{Solutions with a twist potential}
\label{Swtp}

Our next result assumes that $\omega$ is exact and that $X$ has no
zeros. The case with zeros will be covered in Theorems~\ref{T4}
and~\ref{T4n}, while the question of exactness of $\omega$ will be
addressed in Theorem~\ref{T6}; notations and conventions of
Appendix~\ref{Ante} are used.

In the result that follows we assume that  $(\partial M,\ghat)$ is
not conformal to the round sphere.  That last case is covered
by~\cite{AndDahl} when $M$ is spin,  and by~\cite{mand2}
or~\cite{Qing:uniqueness}, (together with~\cite{ChristLohkamp}),
regardless of the existence of a spin structure; in those works it
is shown that $(M,g)$ is then the hyperbolic space. In our context
a simple proof can be given assuming non-degeneracy, for then
every continuous isometry descends to the interior, and the result
follows by ODE methods.

\begin{Theorem}\label{T3} Let $(M,g)$ be Einstein, assume that $(\partial M,\ghat)$ is
not conformal to the round sphere, and suppose that $\hat X(0)$ is
hypersurface-orthogonal on $\partial M$. Assume further that there
exists on $\zSigma$ a $(n-4)$--form $\tau$ such that
\begin{equation} \label{e4}
\omega  = d\tau \;.
\end{equation} If $X$ has no zeros, then
$X$ is hypersurface-orthogonal on $M$.
\end{Theorem}

Both the $(n-4)$--form $\tau$ of \eq{e4}, as well as its
$M$--counterpart $\hat \tau=\pi_\Sigma^*\tau$, will be referred to
as \emph{the twist potentials}.

\medskip

 \proof
We use the notation of Appendix~\ref{Ante} throughout. Let $r$ be
the coordinate of \eq{kill0}. By Remark~\ref{Re2} the function $r$
passes to the quotient $\Sigma=M/S^1$, and by an abuse of notation
we shall use the same letter for the resulting function. For
$\rho>0$ set
$$\Sigma(\rho)=\Sigma\setminus\left(\{r< \rho\}\cup\{p:d(p,\sing)<\rho\}\right)\subset \zSigma\;.$$
By \eq{nke16} we have \bel{nke16a}d (u^{-3} *_{g_\Sigma} \omega) =
0\;.\ee Taking the exterior product of this equation with $\tau$
and integrating over $\Sigma(\rho)$ one has \bean 0 & = &
(-1)^{(n-4)}\int_{\Sigma (\rho)} \tau \wedge d (u^{-3}
*_{g_\Sigma} \omega) \\\nonumber & = & \int_{\partial\Sigma
(\rho)} u^{-3} \tau \wedge *_{g_\Sigma} \omega - \int_{\Sigma
(\rho)} u^{-3} d\tau \wedge *_{g_\Sigma} \omega\\& = &
\int_{\partial\Sigma (\rho)} u^{-3} \tau \wedge *_{g_\Sigma}
\omega - \int_{\Sigma (\rho)} u^{-3} |\omega|_{g_\Sigma}^2
*_{g_\Sigma}1\;. \eeal{nke18a} The idea is to show that the
boundary integral above vanishes when passing with $\rho$ to zero,
yielding $\omega=0$, as desired.

For $\rho$ small enough $\partial\Sigma(\rho)$ is a finite union
of smooth submanifolds of $\zSigma$ of co-dimension one. The
simplest case is $\zSigma=\Sigma$, this occurs when $X$ has no
zeros and all orbits are of principal type, so that
$\sing=\emptyset$ and $\partial\Sigma(\rho)$ equals
$$\partial_\infty\Sigma(\rho):=\{r=\rho\}\;.$$ In general $\partial\Sigma(\rho)$
might have further components of the form
$$\psings(\rho):=\{p:d(p,\sings)=\rho\}$$ and also
$$\psingt(\rho):=\{p:d(p,\singt)=\rho\}\;.$$
The latter are, however, excluded by our current hypothesis that
$X$ has no zeros on $M$.

Lemma~\ref{Lkil} and the definition \eq{e4} of $\tau$
 give\bean (n-3)\partial_{[r} \tau_{A_1\ldots
A_{n-4}]}&=&\omega_{rA_1\ldots A_{n-4}}
\\ & =& O(r)\;,\eeal{nke0d}
where square brackets over a set of indices denote complete
anti-symmetrisation with an appropriate combinatorial factor
($1/((n-3)!)$ in the current case). In dimension four this gives
\bean \partial_{r} \tau & =& O(r)\;,\eeal{nke0da} while in higher
dimensions one obtains \beaa
\partial_{[r} \tau_{A_1\ldots A_{n-4}]}= \partial_{r}
\tau_{A_1\ldots A_{n-4}} + (-1)^{n-4}\partial_{[A_1}
\tau_{A_2\ldots A_{n-4}]r}& =& O(r)\;.\eeaa 
By integration we are led to \bea \tau_{A_1\ldots A_{n-4}}&
=&\sigma_{A_1\ldots A_{n-4}}+
O(1)\;,\eeal{nke0g} where $\sigma_{A_1\ldots A_{n-4}}=0$ in
dimension four, and
$$\sigma_{A_1\ldots
A_{n-4}}:= -\int_{r_0}^r(-1)^{n-4}\partial_{[A_1} \tau_{A_2\ldots
A_{n-4}]r} dr $$ otherwise. Let us use the symbol $\tilde d$ to
denote the exterior differential on $\partial_\infty\Sigma(\rho)$,
at fixed $\rho$. Then
\beaa\sigma&:=& \frac 1 {(n-4)!}\sigma_{A_1\ldots
A_{n-4}}dx^{A_1}\wedge\ldots\wedge dx^{A_{n-4}}
\\ & = &\tilde d \left[-\frac 1 {(n-4)!}\left(\int_{r_0}^r(-1)^{n-4}\tau_{A_2\ldots
A_{n-4}r} dr\right)dx^{A_2}\wedge\ldots\wedge dx^{A_{n-4}}\right]
=: \tilde d \tilde \sigma \;.\eeaa We note that
$$\int_{\partial_\infty\Sigma (\rho)} u^{-3} \tilde d \tilde\sigma   \wedge
*_{g_\Sigma} \omega = \int_{\partial_\infty\Sigma (\rho)} u^{-3}
 d \tilde\sigma   \wedge *_{g_\Sigma} \omega = \int_{\partial_\infty\Sigma (\rho)}
  \tilde\sigma   \wedge d(u^{-3} *_{g_\Sigma} \omega) = 0\;, $$
so that \Eqsone{nke0b} and \eq{nke0g} imply
 \bean
\int_{\partial_\infty\Sigma (\rho)} u^{-3} \tau \wedge
*_{g_\Sigma} \omega & =& \int_{\partial_\infty\Sigma (\rho)}
u^{-3} \tau \wedge\lambda = O(\rho^{n-1})\;,
\eeal{nke0e} which tends to zero as $\rho$ tends to zero. If
$\Sigma=\zSigma$ we are done.

Since we have assumed that $X$ has no zeros, it only remains to
analyse the boundary integral around the $S^1$ orbits with a
non-trivial isotropy group. By point 1 of Proposition~\ref{Lorb}
below such orbits necessarily form a lower-dimensional subset of
$\Sigma$, with $u$ being uniformly bounded in a neighborhood
thereof. We are thus integrating a bounded $(n-2)$--form over a
submanifold, the $(n-2)$--area of which shrinks to zero as $\rho$
tends to zero, which leads to a vanishing contribution in
\eq{nke18a} in the limit. \qed

We wish, next, to prove an equivalent of Theorem~\ref{T3} that
allows zeros of $X$. The proof will again proceed via the identity
\eq{nke18a}, except that we will have now a supplementary
contribution from $\psingt(\rho)$.
 Let $\hF$ be the curvature of
the $U(1)$--principal bundle of unit normals to $\mtwo$, obtained
from the $U(1)$--connection $\gamma_a dx^a$ defined in \eq{afp4a};
in local coordinates, \bel{me16.1} F:=d(\gamma_a dx^a)\;, \quad
\hF:= \pi_\Sigma^* F\;. \ee We shall use the notation and
terminology of Appendix~\ref{Asofp}. We have:

\begin{Theorem}\label{T4} Under the remaining hypotheses of
Theorem~\ref{T3}, assume instead that $\cup_\ell\ml\ne\emptyset$
and that \bel{me8.0a}\int_{\mtwo} \htau\wedge \hF -\frac{2\pi}
{\kappa_1\kappa_2}\int_ {\mfour} \htau =0\;. \ee Then
\bel{me8.1}\cup_{\ell\ge 2}\ml=\emptyset\;, \ee and the
conclusions of Theorem~\ref{T3} hold.
\end{Theorem}

 \proof By Proposition~\ref{Lorb} the set
$\singt$ is the projection by $\pi_\Sigma$ of a disjoint union of
smooth, non-intersecting, submanifolds of dimension $n-2\ell$,
$0\le \ell\le n/2$. It is thus sufficient to consider each such
manifold separately. Consider, then, a connected component of
$\psingt(\rho)$ which is a projection of a connected component of
$\ml$ for some $\ell$, and suppose that
$$\psings(\rho)\cap \psingt(\rho)=\emptyset$$
for $\rho$ small enough. (Proposition~\ref{PB1} shows that this
occurs precisely for those components of $\psingt$ for which the
associated $\ml$'s have all $\kappa_i$'s equal to one.)

 Consider,
first, the case $\ell=1$; using \eq{af[7c}, \eq{nke8} and
\eq{nke0.6} we find (recall that $\hat \lambda$ can be identified
with $\lambda$ in the adapted coordinate system used)
\bean \lim_{\rho\to0}\int_{\psingt(\rho)} u^{-3} \tau \wedge
*_{g_\Sigma}\omega &=&  \int_ {\stwo} \tau\wedge F\\
&=& 
\int_ {\mtwo} \htau\wedge \hF \;.\eeal{me16} In dimension $n$
equal to four the last term in \eq{me16} is the value of $\tau$ at
the connected component of $\mtwo$ under consideration multiplied
by the Euler class of the principal $U(1)$--bundle of unit vectors
normal to $\mtwo$.  Regardless of the dimension $n$, we have:

\begin{Proposition}
\label{R4}  When the normal bundle of $\mtwo$ is trivial the first
integral in \eq{me8.0a} vanishes.
\end{Proposition}
\proof If the normal bundle of $\mtwo$ is trivial, then
$\gamma_adx^a$ is defined globally
 on $\mtwo$, and the integrand in \eq{me16} integrates out to zero:
\bel{me16.2} (-1)^{n-4} \int_ {\selltwo} \tau\wedge d(\gamma_a dx^a)
= \int_ {\selltwo} \left(d\left(\tau\wedge \gamma_a dx^a\right) -
d\tau\wedge \gamma_adx^a\right) = 0\;;\ee recall that $d\htau=0$
on $\ml$. (Strictly speaking, for $\ell\ge 2$ one should do the
above calculation on $\psingt(\rho)$ and then pass to the limit,
since $\psingt$ does not have a differentiable manifold structure
in general for $\ell\ge 2$ --- while $\tau$ extends by continuity
to $\ml$ in the coordinates of Appendix~\ref{Asofp}, the exterior
derivative $d\tau$ of $\tau$ might not be defined there). \qed

Returning to the proof of Theorem~\ref{T4} suppose, next, that
$\ell\ge 2$.
 In the coordinates $(\rho_1,(\rho_i,\psi_i)_{i=2,\ell-1},x^a)$ of \eq{afp17} the
boundary integrand in \eq{nke18a} is of order of $\rho^{-1}$ while
$$\psings(\rho)=\{\rho_i\ge 0\;,\
\rho_1^2+\ldots+\rho_\ell^2=\rho^2\;,\ \psi_i\in [0,2\pi]\;,\
x^a\in \ml\}$$ has (coordinate Lebesgue) measure
$O(\rho^{\ell-1})$, hence \bel{me14} \int_{\psingt(\rho)} u^{-3}
\tau \wedge *_{g_\Sigma}\omega = O(\rho^{\ell-2})\;. \ee The
simplest case is then $\ell\ge 3$, which immediately gives zero
contribution in the limit. \Eq{me14} also shows that the $\ml$'s
with $\ell=2$ give a finite contribution as $\rho$ tends to zero.
Clearly, the only terms that might give a non-zero contribution in
the limit are those which arise from the second line of
\eq{afp17}. If the dimension of $M$ is four then the second term
there does not occur. The first term looks like a total divergence
so one is tempted to conclude that it gives a zero contribution
when integrated upon. This is, however, deceptive, because the
coordinate system used there is singular at the set $\rho_1=0$,
and around each connected component of $\sfour$ from 
\eq{afp17} one finds \bean \lim_{\rho\to0}\int_{\psingt(\rho)}
u^{-3} \tau \wedge *_{g_\Sigma}\omega & = & -\frac{2\pi}
{\kappa_1\kappa_2} \int_ {\sfour} \tau \\ & = & -\frac{2\pi}
{\kappa_1\kappa_2}\int_ {\mfour} \htau \;;\eeal{me16n0} the
$1/\kappa_1=\kappa_1=\pm1$ factor arises from a change of
orientation. In dimension  four each connected component of
$\mfour$ is a point, and the integral here is understood as the
value of $\htau$ at the point under consideration; \eq{me16n0}
gives the contribution from the first term in the second line of
\eq{afp17} for all $n$. It can be checked that  for $n>4$, the
second term there gives a vanishing contribution in the limit, so
that \eq{me16n0} holds for all dimensions. (Strictly speaking, at
this stage $\kappa_2=1$ in the formula above, as we have assumed
that all the nearby orbits have period equal either $0$ or $2\pi$.
However, as shown below, the above formula also gives the boundary
contribution around the $\sfour$'s in general.)

 Clearly \eq{me16n0}
depends only upon the $\sfour$--homology class of the restriction
$\ztau$ of $\tau$ to $\sfour$.

 Let us show how to reduce
the general case to the previous one. As explained in
Appendix~\ref{Asofp}, in the coordinate patch $\mcU_p$ defined
there the surface $\psings(\rho)$ takes the form
$$\psings(\rho)\cap \mcU_p=\left\{\sum_{i= i_2}^\ell\rho_i^2=\rho\right\}\;.$$
We can deform those surfaces to
$$\{\sum_{i= 1}^\ell\rho_i^2=\rho\}$$
using the family of surfaces
$$\partial\Sigma(\rho,\delta):=\underbrace{\left\{\sum_{i= 1}^\ell\rho_i^2=\rho,\sum_{i= i_2}^\ell
\rho_i^2\ge\delta \right\}}_{\partial\Sigma_1(\rho,\delta)} \cup
\underbrace{\left\{\sum_{i= 1}^\ell\rho_i^2\ge\rho,\sum_{i=
i_2}^\ell\rho_i^2=\delta \right\}}_{\partial\Sigma_2(\rho,\delta)}
\;, $$ with $0\le\delta\le \rho$. At fixed $\rho$, on
$\partial\Sigma(\rho,\delta)$ the integrand is uniformly bounded
while $\partial\Sigma(\rho,\delta)$ shrinks to a lower-dimensional
object as $\delta$ tends to zero, therefore
$$
\lim_{\delta\to 0}\int_{\partial\Sigma_2(\rho,\delta)} u^{-3} \tau
\wedge *_{g_\Sigma}\omega = 0\;.
$$
This reduces the problem of calculating the limit, as $\rho$ goes
to zero, of the integral of $u^{-3} \tau \wedge
*_{g_\Sigma}\omega$ over $\psings(\rho)\cap \mcU_p$, to that of
calculating
$$
\lim_{\rho\to 0}\int_{\partial\Sigma_1(\rho,0)} u^{-3} \tau \wedge
*_{g_\Sigma}\omega \;.
$$
But this is an integral already considered under the assumption
that $\psings$ does not meet $\psingt$ in $\mcU_p$, and an
identical analysis applies.

Those components of $\sings$ which do not meet $\singt$, or which
lie away from the $\mcU_p$'s, are handled as in the proof of
Theorem~\ref{T3}.
 Finally, \eq{me8.1} is a rephrasing of
Proposition~\ref{PA1}. \qed

There are various ways to ensure that \eq{me8.0a} holds: Suppose,
for instance, that we are in dimension four, then $\htau$ is a
function on $M$, defined up to a constant; further $\htau$ is
constant on any connected component of $\ml$. In this case, when
$\cup_{\ell\le 2}\ml$ is connected, we can choose $\tau$ to be
zero on $\cup_{\ell\le 2}\ml$, obtaining a vanishing contribution
from $\cup_{\ell\le 2}\ml$.  Another possibility is to assume that
the bundle of unit normals to $\mtwo$ is trivial, see
Proposition~\ref{R4}. If, moreover, $\mfour$ is connected (which
will certainly be the case if it is empty), then we can choose
again the constant of integration appropriately to achieve the
desired equality. One can clearly assume various combinations of
the hypotheses above. As a special case, we have
obtained:

\begin{Theorem}\label{T4n} Under the remaining hypotheses of
Theorem~\ref{T3}, assume instead that
\bel{me8.0}\begin{cases}\cup_{\ell\le 2}\ml \ \mbox{ is connected,} & n=4 \;, \mbox{or}
\\ \mbox{the normal bundle to $\mtwo$ is trivial and}\ \mfour \ \mbox{is connected,} & n= 4 \;,  \mbox{or}
\\ \mbox{the normal bundle to $\mtwo$ is trivial and}\ \mfour=\emptyset \;, & n\ge 4 \;,  \mbox{or}\\
\cup_{\ell\le 2}\ml=\emptyset\;, & n\ge 4\;. \end{cases}  \ee Then
the conclusions of Theorem~\ref{T4} hold.
\end{Theorem}

  \qed

  We note that the hypotheses of Theorem~\ref{T4n} are stable
  under perturbations of the metric.

\subsection{Existence of the twist potential} \label{Stwist}

Let us briefly turn our attention  to  the question of existence
of the twist potential; this will be obviously  the case when
$H_{n-3}(M)$ is trivial. Such a hypothesis, however, excludes most
situations of interest from a Lorentzian point of view if $n=4$.
An alternative possibility is triviality of $H_{n-3}(\zSigma)$ ---
this covers, in particular, all Lorentzian space-times without
black hole regions, with $M=\Sigma\times S^1$, and with trivial
$H_{n-3}(\Sigma)$.

 In dimension four, a further family of examples
can be obtained as follows: It follows from Lemma~\ref{Lkil} that,
in the coordinate system of \eq{kill0}, the one forms
$$\homega_{A}dx^A \ \mbox{ and }\  \omega_{A}dx^A $$ extend by continuity to
closed one-forms $\homega_0$ on $\partial M$, and $\omega_0$ on
$\partial \hyp$. Clearly a necessary condition for exactness of
$\homega$ is exactness of $\homega_0$. Under some conditions this
can be shown to be sufficient:
\begin{Theorem}\label{T6} In dimension $n=4$, suppose that $X$ has no zeros, then the twist
potentials $\tau$ and $\htau$ exist under either of the following
conditions:
\begin{enumerate} \item\label{T6p1} There exists a
function $\htau_0$ on $\partial M$ such that $d\htau_0=\homega_0$,
and there are no non-trivial $L^2$ sections $\varphi$ of
$\Lambda^1(M) $ which are solutions of the equations \bel{ke7}
d\varphi = d*_g\varphi=0\;.\ee \item\label{T6p2} There are no
$S^1$ orbits with nontrivial isotropy, there exists a function
$\tau_0$ on $\partial M$ such that $d\tau_0=\omega_0$, and there
are no non-trivial $L^2$ sections $\varphi$ of
$\Lambda^1(\zSigma)$ which are solutions of the equations
\bel{ke7a} d\varphi = d*_{g_\Sigma}\varphi=0\;.\ee
\end{enumerate}
\end{Theorem}

\begin{Remark} Wang~\cite[Theorem~3.1]{MR1879811} gives a condition under which
the $L^2$--cohomology condition above will hold; in particular it
follows from the work of Lee~\cite{MR96h:58176} that the
$L^2$--cohomology condition  will be satisfied when the Yamabe
invariant of the boundary metric on $\partial M$ or on $\partial
\hyp$ is positive.
\end{Remark}

\proof We first note that existence of $\tau$ and $\htau$ are
equivalent, by projecting down or lifting. In order to prove point
\ref{T6p1} consider the equation \bel{se0a} \nabla_i\nabla^i \ttau
= \frac 4 u \homega_i\nabla^iu\;.\ee Lemma~\ref{Lkil}  and the
calculations there show that $ u^{-1}\homega_i\nabla^iu= O(r^2)$,
so that by point (ii) of Theorem~7.2.1 of \cite{AndChDiss},
together with the Remark (i) following that theorem, there exists
a function
$$\ttau=\htau_0+O(r^2)$$ which solves \eq{se0a}. \Eq{ke2} shows that the one-form
$$\varphi:=\homega_idx^i-d\ttau$$ solves \eq{ke7}. Further we
have, in the coordinates of Appendix~\ref{Aext},
$$\partial_i \rfloor \varphi = O(r) \quad \Big(\mbox{equivalently,} \quad |\varphi|^2_g = O(r^4)\Big) \;,$$
which implies that $\varphi\in L^2$. The vanishing of $\varphi$
follows from our hypothesis of the vanishing of the first
$L^2$--cohomology class of $M$, hence $\homega=d\ttau$. Point
\ref{T6p2} is established in
a similar way, using \eq{nke16} instead of \eq{ke2}. 
 \qed

\appendix
\section{Extensions of conformal isometries from $\partial M$ to
$M$}\label{Aext}

Let $Y$ be a conformal Killing vector field of $\partial M$.
Suppose, first, that $(\partial M,\ghat)$ is a round sphere; as
discussed at the beginning of Section~\ref{Swtp}, the pair $(M,g)$
is then the hyperbolic space. Consider, next, a $\partial M$ which
is \emph{not} a round $(n-1)$--dimensional sphere, then the
Lelong-Ferrand -- Obata theorem shows that we can choose a
representative $\hat g(0)$ of the conformal class $[\ghat]$ so
that $Y$ is a Killing vector thereof. We start with a study of the
Killing equation in a neighborhood of $\partial M$. It is well
known that there exists a defining function $r$ such that the
metric $g$ takes the form
\bel{kill0} g=r^{-2}\overline{g}=r^{-2}(dr^2+\hat{g}(r))\;,\qquad
\hat{g}(r)(\partial_r,\cdot)=0\;. \ee on
$[0,\epsilon]\times\partial M$. Let $(x^{2},...,x^{{n}})$ be a
local coordinate system on $\partial M$. We will work in the
coordinate system $(x1 = r, x2, ... x^n)$, and denote by $r$ the
index relative to the first coordinates. We will take upper case
Latin letters for the indices relative to the remaining
coordinates, and lower case Latin letters for the indices relative
to any component. With that convention, the Christoffel symbols of
$g$ read \beal{Chr1} &\Gamma^r_{rr}=-r^{-1},\;\;
\Gamma^A_{rr}=\Gamma^r_{Ar}=0,\;\;\Gamma^r_{AB}=r^{-1}\hat{g}_{AB}(r)-\frac{1}{2}\hat{g}'_{AB}(r),
&
\\ & \Gamma^C_{rA}=-r^{-1}\delta^C_A+\frac{1}{2}\hat{g}^{CD}(r)\hat{g}'_{DA}(r),\;\;
\Gamma^C_{AB}=\hat{\Gamma}^{C}_{AB}(r)\;. \eeal{Chr2} Here $f'$
denotes the derivative of a function $f$ with respect to $r$. The
Killing equations, \be\label{kill} \nabla_iX_j+\nabla_jX_i=0, \ee
written out in detail, read \be\label{killrad}
\partial_rX_r+r^{-1}X_r=0,
\ee \be\label{killmix}
\partial_rX_A+\partial_A
X_r+2r^{-1}X_A-\hat{g}^{CD}(r)\hat{g}'_{DA}(r)X_C=0,
\ee\be\label{killtan}
\partial_AX_B+\partial_BX_A-2\hat{\Gamma}^C_{AB}(r)X_C+(\hat{g}'_{AB}(r)-2r^{-1}\hat{g}_{AB}(r))X_r=0.
\ee From \eq{killrad} there exists a function $\alpha$ such that
$$
X_r=\frac{\alpha}{r},\qquad \partial_r\alpha=0\;,
$$
and, if we define $\hat{X}_A=r^{2}X_A,$ then \eq{killmix} and
\eq{killtan} become \be\label{killmix2}
\partial_r\hat{X}_A+r\partial_A
\alpha-\hat{g}^{CD}(r)\hat{g}'_{DA}(r)\hat{X}_C=0,
\ee\be\label{killtan2}
\partial_A\hat{X}_B+\partial_B\hat{X}_A-2\hat{\Gamma}^C_{AB}(r)\hat{X}_C+(r\hat{g}'_{AB}(r)-2\hat{g}_{AB}(r))\alpha=0.
\ee From \eq{killtan2}, $\hat{X}(0)$ is a Killing vector field of
the boundary if and only if $\alpha\equiv 0\Leftrightarrow
X_r\equiv 0$. If that is the case then \eq{killmix2} and
\eq{killtan2} take the form \be\label{killmix3}
\partial_r\hat{X}_A-\hat{g}^{CD}(r)\hat{g}'_{DA}(r)\hat{X}_C=0,
\ee\be\label{killtan3}
\partial_A\hat{X}_B+\partial_B\hat{X}_A-2\hat{\Gamma}^C_{AB}(r)\hat{X}_C=0.
\ee \Eq{killmix3} has the unique solution \bel{kill2}
\hat{X}_A(r)=\hat{g}_{AC}(r)\hat X^C(0)\;, \qquad \hat
X^C(0):=\hat{g}^{CB}(0)\hat{X}_B(0). \ee We use now the Taylor
development
$$\hat{g}(r)=\hat{g}(0)+O(r^p),$$ where $p=1$ in general and $p=2$
if $g$ is Einstein~\cite{FG}. This yields
$\hat{X}(r)=\hat{X}(0)+O(r^p)$ and
$\hat{\Gamma}(r)=\hat{\Gamma}(0)+O(r^p)$, thus $\hat{X}$ given by
\eq{kill2} is an approximate solution of \eq{killtan3} modulo
$O(r^p)$. Finally the 1-form
\bel{e0.0}
X_\infty:=0\,dr+r^{-2}(\hat{X}_{2}(r)d{x^2}+\ldots+\hat{X}_{{n}}(r)d{x^{n}})
\ee is an approximate solution of \eq{kill}, with error -- in the
above coordinates -- $O(r^{p-2})$.

We wish to show  that, under reasonably mild conditions, conformal
isometries of $[\ghat]$ extend to isometries of $\riemg$:

\begin{Proposition}\label{killglob} Let $(M,g)$ be an asymptotically hyperbolic
Einstein manifold and suppose that the operator
$$\Delta_L+2(n-1)$$ acting on symmetric two-covariant tensors has
no $L^2$ kernel. Then every Killing vector field $\hat{X}(0)$ on
$\partial M$ extends in a unique way to a Killing vector field $X$
on $M$ such that \eq{e1} holds.
\end{Proposition}

 For
$\hat{X}(0)$, a (one form associated to a) Killing vector field on
$\partial M$, consider the boundary value problem \bea &&\Delta_g
X_i = -\Ric(g)_{i}{}^{j} X_j\;,\label{e1.0}
\\ &&X-X_\infty \in C^{2,\alpha}_p(M,T^*M)\;, \eeal{e1}
with $p$ as defined in the paragraph before \eq{e0.0}. We have
\begin{Proposition}\label{P1} Let $(M,g)$ be an asymptotically hyperbolic
 manifold with $\Ric (g)<0$. Then the problem
\eq{e1.0}-\eq{e1} always has a unique solution.
\end{Proposition}

\proof  From Mazzeo~\cite{Mazzeo:hodge} (see also
\cite[Lemma~7.2]{Lee:fredholm}) the indicial radius of the
Laplace-Beltrami operator $dd^*+d^*d$  on one-forms, equal to
$\nabla_g^*\nabla_g+\Ric(g)$ on those, is $(n-1)/2-1$. Corollary
7.4 in~\cite{Lee:fredholm} implies then that the indicial radius
of the operator \bel{Pop}P=\nabla_g^*\nabla_g-\Ric(g)\ee on
one-forms is $ \sqrt{[(n-1)/2-1]^2+2(n-1)}=(n-1)/2+1$. An
integration by parts shows that there are no $C^2$ compactly
supported solutions of \eq{e1.0}:
$$0\le \int -\Ric(g){}_{ij}X^iX^j = \int X_i\Delta_g X^i =- \int |\nabla X|^2
\le 0$$  (recall $\Ric(g)<0$). Elliptic regularity, completeness
of $M$ together with  density results (\emph{cf.,
e.g.,}\/~\cite{Aubin76}) imply that $P$ has trivial $L^2$ kernel,
and~\cite[Theorem C]{Lee:fredholm} establishes that, in the
notations of~\cite{Lee:fredholm},  $P$ is an isomorphism from
$C^{k,\alpha}_\delta(M;T^1)\equiv C^{k,\alpha}_\delta(M;T^*M)$ to
$ C^{k-2,\alpha}_\delta(M;T^*M)$ for all $\delta$ such that
$|\delta-(n-1)/2|<(n-1)/2+1\Leftrightarrow-1<\delta<n$. (In the
case of vector fields, the space
 $C^{k-2,\alpha}_{-1}(M;T^*M)$ corresponds to  $O(r^{-2})$ behavior in the coordinates of \eq{kill0}.)
 Let $\chi$ be a smooth function on $M$
equal to $1$ on $\{0\leq r\leq\epsilon/3\}$, and equal to $0$ for
$r\ge 2\epsilon/3$; define the $1-$form
$$
Y=\chi X_\infty\;,
$$
with $X_\infty$ defined in \eq{e0.0}. Then $Y$ is an approximate
solution to the Killing equation \eq{kill} modulo $O(r^{p-2})$.
(We emphasise that \eq{killrad} and \eq{killmix} are satisfied identically near
the boundary, so that the fall-off of the error term is dictated by
a possible error in \eq{killtan}.)
This implies $PY=O(r^{p-1})\in C^{k-2,\alpha}_{p}(M;T^*M)$ (see,
\emph{e.g.}, the proof of \cite[Lemma~3.7]{Lee:fredholm} for that
last property). Thus there exists a unique solution $Z\in
C^{k,\alpha}_{p}(M;T^*M)$ to the equation
$$
PZ=-PY\;.
$$
We set $X=Y+Z$; uniqueness is obvious from what has been said
above. \qed\

{\noindent \sc Proof of Proposition~\ref{killglob}:} If we denote
by $B(h)=-\tr_g \nabla h+\frac{1}{2}\nabla \tr_g(h)$, then the
linearisation of the Einstein operator at the Einstein metric $g$
is~\cite[Theorem~1.174]{Besse}
$$
D\Ein(g)=\frac{1}{2}(\Delta_L+2(n-1))-\Div^*B.
$$
Let $X$ be the solution of \eq{e1.0}-\eq{e1}. Then $X$ is in the
kernel of the operator $P$ of \eq{Pop}. A two-line calculation
shows that $$B({\cal L}_{X}g)=P(X)=0\;.$$ Now, if $g$ is an
Einstein metric then ${\cal L}_Xg$ is in the kernel of $D\Ein(g)$
whatever the vector field $X$: if $\phi_t$ denotes the (perhaps
local) flow of $X$, then
$$0=\frac d {dt} (\phi_t^*(\Ein(g)))\Big|_{t=0}= \frac d {dt}
(\Ein(\phi_t^*g)) \Big|_{t=0}= D\Ein(g) {\cal L}_Xg\;.$$ It thus
follows that
$$(\Delta_L+2(n-1)){\cal L}_Xg=0\;.$$ Now,  Theorem C
and Proposition D of~\cite{Lee:fredholm} show that the operator
$\Delta_L+2(n-1)$ is an isomorphism from $
C^{k,\alpha}_\delta(M,S_2)$ to $ C^{k-2,\alpha}_\delta(M,S_2)$ for
all $\delta$ such that
$$|\delta-(n-1)/2|<
(n-1)/2 \quad \Longleftrightarrow \quad 0<\delta<n-1\;.$$ Here we
use the symbol $S_2$ to denote the bundle of symmetric
two-covariant tensors;  in the notation of \cite{Lee:fredholm} the
space $ C^{k,\alpha}_0(M,S_2)$ corresponds to $O(r^{-2})$ behavior
in the coordinates of \eq{kill0}. From what has been said we have
${\cal L}_Xg=O(r^{p-2})$ in local coordinates near the boundary,
which can be written as ${\cal L}_Xg\in C^{k,\alpha}_{p}(M,S_2)$.
Since $p>0$ the isomorphism property gives
$$
{\cal L}_Xg=0.
$$
\qed
\begin{remark}\label{Re2}
As $X$ is a Killing vector field, $\hat{X}$ satisfies
\eq{killmix3} and \eq{killtan3}, in particular the field of
covectors
$\hat{X}_A(r_0)=\hat{g}_{AC}(r_0)\hat{g}^{CB}(0)\hat{X}_B(0)$
satisfies the Killing equations on the hypersurface $\{r=r_0\}$.
This is equivalent to the statement that  $X$ is tangent to the
level sets of $r$ with $\hat X^A(r)=\hat X^A(0)$. Further, in the
coordinate system of \eq{kill0},
 \bel{mainkil} \xi:= g(X,\cdot) =
r^{-2}(\hat{X}_{2}(r)d{x^2}+\ldots+\hat{X}_{{n}}(r)d{x^{n}})\;.
\ee
\end{remark}
%
%
%
%
%
%

 \section{The norm and twist equations}
 \label{Ante} Let
$(M,g)$ be an $n$-dimensional Riemannian or Lorentzian space-time
with a Killing vector field $X$, \bel{ke0.0} \nabla_i X_j +
\nabla_j X_i=0\;.\ee It is well known that \eq{ke0.0} implies the
equation \bel{ke0} \nabla_i\nabla_j X^k = R_{\ell ij}{}^k
X^\ell\;,\ee in particular \bel{ke0.1} \nabla^j\nabla_j X^k = -
\Ric^k{}_j X^j\;.\ee Let us, locally, write the metric in the form
\eq{e2}:
\begin{equation} \label{ne2}
g = \eta u^{2}(d\phi  + \theta )^{2} + g_{\Sigma}\;,\quad
\theta(\partial_\phi)=g_{\Sigma}(\partial_\phi,\cdot)=0\;, \quad
X=\partial_\phi\;,
\end{equation}
where $g_\Sigma$ is the metric induced by $g$ on the distribution
$X^\perp\subset TM$,  and $\eta=\pm 1$ according to whether $X$ is
spacelike ($\eta = 1$) or timelike ($\eta = -1$). The metric
$g_\Sigma$ is the natural metric on the orbit space
$\Sigma$~\cite{Geroch:1971nt}. One can also think of $\Sigma$  as
of any hypersurface transverse to $X$, regardless of the structure
of the flow of $X$; one should then, however, not confuse
$g_\Sigma$ with the metric induced by $g$ on $\Sigma$. We will be
interested in the equations on $\Sigma$; an efficient way of
obtaining those is provided by the projection formalism of
Geroch~\cite{Geroch:1971nt}. We will be working away from the set
of zeros of $g(X,X)$. Let
$$P:TM\to TM$$ denote the orthogonal projection on $X^\perp$, we
will also use the symbol $P$ to denote the obvious extension of
$P$ to other tensor bundles. We note that \beal{ke1a} u&=&
\sqrt{\eta g(X,X)}\;,\eea (which can be used as the definition of
$u$ regardless of the decomposition \eq{ne2}) and we set \bea n&:=
&\frac X u\;.\eeal{ke1n} We then have
$$P(Y)= Y- \eta g(Y,n)n=(\delta^i_j - \eta n^i n_j) Y^j \partial_i\;.$$
If $Y$ and $Z$ are tangent to $\Sigma$, and if $\hat Y$ and $\hat
Z$ are $X$-orthogonal, $X$-invariant lifts of $Y$ and $Z$ to $M$,
then the covariant derivative defined as
$$D_Y Z:= P(\nabla_{\hat Y} \hat Z)$$
is the Levi-Civita covariant derivative of $g_\Sigma$ (see
\cite{Geroch:1971nt}). Let $$\hlambda:= P(u\nabla X)\;,$$ so that
\bel{nke0}\hlambda_{ij}= u \nabla_i X_j + X_i u_j - X_j u_i\;,\ee
where we have written $u_j$ for $\nabla_j u$. The tensor field
$\hat \lambda$ is well defined and smooth away from the set of
zeros of $u$ (at which $u$ might fail to be differentiable).
 One has $X^i\nabla_iu=0$ by
\eq{ke0.0}, and one easily checks that $\hlambda$ is an
anti-symmetric $X$-invariant tensor field on $M$ which annihilates
$X$, and thus defines a two-form $\lambda$ on $\Sigma$ in the
usual way. A convenient way of calculating $\lambda$ in practice
is to introduce \bel{nke0.4}\beta:= u^{-2} \xi \;.\ee With a
little work one finds \bel{nke0.6}\hlambda=
u^3 d\beta \;,\ee which clearly leads to  \bel{nke0.8}
d(u^{-3}\hlambda)=0 \;.\ee It can be seen that $X$ is (locally)
hypersurface orthogonal if and only if $\hlambda$ vanishes.
Indeed, \eq{ne2} shows that the distribution $X^\perp$ is
(locally) integrable if and only if
$$d\phi + \theta = \frac \eta {u^2}g(X,\cdot)$$ is closed; that last
condition is precisely the equation $\hlambda=0$.

Let us derive the equations satisfied by $\hlambda$ and $\lambda$.
Using \eq{ke0.1} we have\bel{nke1}\nabla_k \hlambda_{ij} =u_k
\nabla_i X_j +u R_{skij} X^s + \nabla_k X_i u_j - \nabla_k X_j
u_i+ X_i \nabla_k u_j - X_j \nabla_k u_i\;.\ee
Applying a projection to both sides of \eq{nke1} one finds
\bel{nke2}D_k \lambda_{ij} =\frac{1}{u} u_k \lambda_{ij} +u
P(R_{skij} X^s) + \frac{1}{u}(\lambda_{ki} u_j - \lambda_{kj}
u_i)\;.\ee Projections commute with anti-symmetrisations, so that
the first Bianchi identity implies
$$D_{[k} \lambda_{ij]} =\frac{3}{u}u_{[k} \lambda_{ij]}\;,$$
where square brackets denote complete anti-symmetrisation.
Equivalently, \bel{nke4}d(u^{-3} \lambda)=0\;,\ee where $d$ is
taken on $\zSigma$. This does imply \eq{nke0.8} by pull-back with
$\pi_\Sigma$, but the implication the other way round does not
seem to be evident.

We want to calculate the divergence of $\lambda$. In order to do
that we need to work out the $P(R_{skij} X^s)$ term appearing in
\eq{nke2}; using the fact that $n$ is proportional to $X$ we find
\beaa P(R_{skij} X^s) & = & P( (\delta_j^\ell - n_j
n^\ell)R_{ski\ell} X^s)
\\ & = & P(( R_{skij} - n_j
n^\ell R_{ski\ell} )X^s)
\\ & = & P((\delta_i^m - n_i
n^m)( R_{skmj} - n_j n^\ell R_{skm\ell} )X^s)\\
& = & (R_{skij}- n_i n^mR_{skmj} - n_j n^\ell R_{ski\ell} )X^s\;,
\eeaa where the last equality arises from the fact that all
projections have already been carried out; no projection is needed
in the $k$ index since $n$ is proportional to $X$. Upon a
contraction over $k$ and $i$ in \eq{nke2} the $P(R_{skij} X^s)$
term will thus give a contribution
$$(-R_{sj}- 0 +n_j n^\ell R_{s\ell} )X^s=0$$
if $g$ is Einstein. It follows that, for Einstein metrics $g$,
 \bel{nke4n}D^i
\lambda_{ij} =\frac{1}{u} (u^i \lambda_{ij} + 0
+\underbrace{\lambda_{i}{}^i}_0 u_j - \lambda_{ij} u^i)=0\;.\ee
Equivalently, \bel{nke6} d(*_{g_\Sigma} \lambda)=0\;,\ee with $d$
again taken on $\zSigma$. We define the \emph{twist $n-3$ form
$\omega$ on $\zSigma$} by  the equation \bel{nke8}\omega:=
*_{g_\Sigma} \lambda\;.\ee \Eq{nke6} shows that $\omega$ is
closed, while \eq{nke4} is equivalent to \bel{nke10}d^*(u^{-3}
\omega)=0\;,\ee Let $\hat\omega$ denote the lift of $\omega$ to
$M$,
$$\hat\omega=\pi_\Sigma^* \omega\;,$$
where $\pi_\Sigma$ is the projection from $M$ to $\Sigma$.
Choosing the orientation of $\Sigma$ appropriately one  finds
\bel{nke12} \hat\omega_{\alpha_1\ldots \alpha_{n-3}}=
\epsilon_{\alpha_1\ldots \alpha_{n-3}\alpha\beta\gamma}X^\alpha
\nabla^\beta X^\gamma\quad \Longleftrightarrow\quad \hat\omega =
*_g( \xi\wedge d\xi) \;,\ee where
$$\xi= g(X,\cdot)\;,$$ and where
$\epsilon$ is the volume form on $M$. Since exterior
differentiation commutes with pull-back we have
\bel{nke14}d\hat\omega = d(\pi_\Sigma^* \omega) = \pi_\Sigma^*
(d\omega)=0\;.\ee Summarising, on $\Sigma$ we have \bel{nke16}
 d \omega = d (u^{-3}
*_{g_\Sigma} \omega) = 0\;,\ee while on $M$ it holds that
\bel{nke18} d\hat \omega = d (u^{-3}  \hat \lambda) = 0\;.\ee It
is worthwhile mentioning that so far all the equations were
manifestly signature-independent.

 We
note the following equations for $u$: \bean \nabla_i u &=& \frac
\eta {u} X^j \nabla_i X_j\;,
\\ \nonumber
\nabla^i\nabla_i u &=& \frac \eta {u} (-\eta g(\nabla u,\nabla
u)+\nabla^i X^j \nabla_i X_j - \Ric_{ij}X^iX^j) \\ &=& \frac 1 {u}
( g(\nabla u,\nabla u)+\eta \frac{\hlambda^{ij}\hlambda_{ij}}{u^2}
- {\eta }\Ric_{ij}X^iX^j)\;. \eeal{nke4na} The reader is warned
that the $\hlambda^{ij}\hlambda_{ij}$ term above can sometimes be
negative when $g$ is Lorentzian and $X$ is spacelike; similarly
$g(\nabla u,\nabla u)$ can sometimes be negative for Lorentzian
metrics.

We refer to \cite{Coquereaux:1988ne} for explicit formulae for the
curvature tensor of $g_\Sigma$.

For completeness let us recall how this formalism works in
dimension four: one then sets\beal{ke1} \omega_i&:=&
\epsilon_{ijk\ell} X^j \nabla^k X^\ell\;.\eea Here, as before,
$\epsilon_{ijkl}$ is the volume form,
$$\epsilon_{ijkl}=0,\pm  \sqrt{|\det g_{mn}|}\;,$$
with $\epsilon_{ijkl}$ totally antisymmetric, the sign being $+$
for positive permutations of $1234$.  The form $\omega$ from
\eq{ke1} is actually the form $\hat \omega$ from \eq{nke12}, but
we shall not make a distinction between $\omega$ and $\hat \omega$
anymore. One has $X^i\omega_i=0$ by antisymmetry of
$\epsilon_{ijk\ell}$.
Working away from the set of zeros of $u$, with a little work one
finds \bel{ke2.0} \nabla_i X_j = \frac 2
 u X_{[j}\nabla_{i]}u + \frac {\sigma \eta}{2u^2}
\epsilon_{ijk\ell}\omega^k X^\ell \;,\ee where $\sigma=+1$ in the
Riemannian case, and $\sigma=-1$ in the Lorentzian one. The
simplest way of performing the algebra involved in this equation,
as well as in \eq{ke2} below, is to consider a frame in which $X=
u e^1$, with $\omega$ proportional to $e^2$. Comparing with
\eq{nke0}, one recognises the last term above as $\lambda_{ij}$.
The divergence of $\homega$ can also be computed directly as
follows: \bean \nabla^i \homega_i &=& \epsilon_{ijk\ell}(\nabla^i
X^j \nabla^k X^\ell + X^j R_{m}{}^{ik\ell}X^m)\\&=&
\epsilon_{ijk\ell}\nabla^i X^j \nabla^k X^\ell \nonumber \\ &=&
\frac{4\homega_i} u \nabla^iu \label{ke2}\;. \eea Equivalently,
$$\nabla^i(u^{-4}\homega_i)=0\;.$$
\Eq{nke14} can be rewritten as \bea \nabla_i \homega_j -
\nabla_j\homega_i=D_i \omega_j - D_j\omega_i = 0\;.\eeal{ke3} It
follows that \bean \nabla^k\nabla_k \homega_i
&=&\nabla^k\nabla_i \homega_k= \nabla_i \nabla^k \homega_k + \Ric_{ij}\homega^j \\
& = &4\nabla_i (\frac{\homega_j\nabla^ju}{u})  +
\Ric_{ij}\homega^j \;.\eeal{ke3.1} In the four-dimensional case
the last line of \eq{nke4na} can be rewritten as
\bea \nabla^i\nabla_i u &=&\frac 1 {u} ( g(\nabla u,\nabla
u)+\frac 1 {2u^2}\sigma g(\homega,\homega) - {\eta
}\Ric_{ij}X^iX^j)\;. \eeal{ke4}

\section{The structure of the orbit space near fixed points}
\label{Asofp}

 In order to analyse the contribution to \eq{nke18a}
arising from the integral over $\psingt(\rho)$, we need to recall
some results about the structure of $\psingt$. Since $\nabla_i
X_j$ is antisymmetric, for every $p\in M$ there exists an ON basis
of $T_pM$ in which $\nabla_i X_j$ is block-diagonal, with $\ell$
non-zero anti-symmetric two-by-two blocks eventually followed by a
block of zeros; such a basis will be referred to as \emph{a basis
adapted to $\nabla X$}. It follows that the dimension of the set
 \bel{me0} \kerp:= \{Y\in T_pM: \nabla_Y X=0\}\ee is necessarily a number of the form $n-2\ell$
for some $0\le \ell\le n/2$. For such $\ell$'s we define
 \bea\label{me1} \ml &:=&\{p\in M: X(p)=0\;,\ \mathrm{dim}(\kerp)= n-2\ell\} \;.
\eea For $p\in M$ let $\mathrm{Iso}(p)$ denote the isotropy group
of $p$. We set \bea\label{me1.a} \miso &:=&\{p\in M:\ X(p)\ne
0\;,\ \mathrm{Iso}(p)\ne \mathrm{Id}\}\;. \eea For $p\in \miso$
let $\tp\in \{2\pi/n\}_{n\in\N^*}$ denote the period of the orbit
of $X$ through $p$, set \bea \nonumber \kerpi &:=& \{Y\in T_pM:
(\phi_{\tp})_*Y=Y\}\;,\\
 \mil& :=&\{p\in \miso: \ \mathrm{dim}(\kerpi)= \ell\}\;.
\eeal{me1.b} The following is certainly well known; we give the
proof for completeness, because some elements of the argument will
be needed in our further analysis:
\begin{Proposition}\label{Lorb}
\begin{enumerate}
\item The $\mil$'s are smooth, totally geodesic
$\ell$--dimensional submanifolds of $M$.\item The $\ml$'s are
smooth, closed, totally geodesic $(n-2\ell)$--dimensional
submanifolds of $M$. In particular
$$\mi\cap \mj=\emptyset \ \mbox{ for } \ i\ne j\;.$$
\end{enumerate}
\end{Proposition}

\proof 1.  Let $p\in \mil$ and let $\gamma:[0,s_p)\to M$ be a
maximally extended distance-parameterised geodesic such that
$\gamma(0)=p$ and $\dot\gamma(0)=Y$ for some $Y \in T_pM$. If
$Y\in \kerpi$, then $\phi_{\tp}\circ\gamma:[0,s_p)\to M$ is again
a maximally extended geodesic through $p$ with  tangent vector
$Y$, which implies $\phi_{\tp}\circ\gamma(s)=\gamma(s)$ for all
$s\in [0,s_p)$. It follows that the group orbit through
$\gamma(s)$ has period $\tp$ for  $s$ small enough. Clearly if
$Y\not\in \kerpi$ then we will have
$\phi_{\tp}\circ\gamma(s)\ne\gamma(s)$ again for $s$ small enough,
and the result follows.

2.  For $s\in \R$ let $\phi_s$ denote the action of $S^1$ on $M$,
with $s$ normalised so that $2\pi$ is the smallest strictly
positive number $s_*$ for which $\phi_{s_*}$ is the identity on
$M$. At points at which $X$ vanishes we have, for any vector field
$Y$,
$${\mycal L}_X Y =[X,Y]=\nabla_XY-\nabla_YX= - \nabla_Y X\;,$$
so that \bel{me2} \frac {d(\phi_s{}_* Y)}{ds}=0 \ \mbox{ for } \ Y
\in \kerp\;.\ee Consider any maximally extended affinely
parameterised geodesic $\gamma:I\to M$ with $\gamma(0)=p$, and
with the tangent $\dot \gamma(0)\in \kerp$. Then $\phi_s(\gamma)$
is again a maximally extended affinely parameterised geodesic
through $p$. Further, \bel{me3} \frac {d(\phi_s{}_* \dot \gamma
(0))}{ds }=0\ee by \eq{me2}, which shows that
$$\forall s \quad \frac{d(\phi_s\circ\gamma)(t)}{dt}\Big|_{t=0}=\dot \gamma (0)\;.$$
This implies of course that  $\phi_s(\gamma(t))=\gamma(t)$ for all
$t\in M$ and $s\in\R$, so  that all points on $\gamma$ are fixed
points of $\phi_s$. Hence
$$\exp_p(\kerp)\subset\cup_\ell \ml\;.$$
If we move away from $p$ in a direction which is not in $\kerp$
then $X$ immediately becomes non-zero, which shows that there
exists a neighborhood of $p$ such that $\exp_p(\kerp)$ coincides
with $\ml$ there, and the fact that $\ml$ is a smooth embedded
totally geodesic submanifold follows.

To prove closedness of $\ml$ consider normal coordinates centred
at $p$. After performing a rotation if necessary we may  suppose
that the basis $\{\partial_i\}$ is adapted to $\nabla X$ at $p$,
so that there exist real numbers $\kappa_i=\kappa_i(p)$ such that
at $p$ we have \bel{me4}\frac 12 \nabla_iX_j\, dx^i\wedge dx^j =
\sum_{i=1}^{\ell}\kappa_i \, dx^{2i-1}\wedge dx^{2i}\;.\ee
Closedness of $\ml$ is clearly equivalent to the statement that
the $|\kappa_i|$'s are uniformly bounded away from zero on each of
the $\ml$'s.
It is shown below that the $\kappa_i$'s are integers, and
continuity of the map $\ml\ni p \to \{\kappa_i(p)\}\in \R^\ell$
proves the result. \qed

In order to continue our  analysis of the geometry near fixed
points let $p\in\ml$, with $\ml$ as in \eq{me1}, let $x^a$ denote
any local coordinates on $\ml$ on a coordinate $\ml$--neighborhood
$\mcO_p\subset \ml$ of $p$, and for $q\in \mcO_p$ let $x^A$ denote
geodesic coordinates on $\exp_q\{(T_q\ml)^\perp\}$.  Passing to a
subset of $\mcO_p$ if necessary one obtains thus a coordinate
system $(x^i)=(x^A,x^a)$, with $A=1,\ldots,2\ell$, on an
$M$--neighborhood $\mcU_p\subset M$ of $p$ diffeomorphic to
$\mcO_p\times B_{2\ell}(r)$, where $B_{2\ell}(r)$ is  a ball of
radius $r$ centred at the origin in $\R^{2\ell}$. Since $\ml$ is
compact, it can be covered by a finite number of such coordinate
systems. This leads to the following local form of the metric
\bel{afp0} g= \sum_{i=1}^{2\ell}
(dx^i)^2 + h +\sum_{A,a}O(\rho)dx^Adx^a +
\sum_{A,B}O(\rho^2)dx^Adx^B +\sum_{a,b}O(\rho^2)dx^adx^b   \;,\ee
with $h$
--- the metric induced by $g$ on $\ml$, where $\rho$ denote the
geodesic distance to $\ml$. The $O(\rho^2)$ character of the $dx^A
dx^B$ error terms is standard;  the $O(\rho^2)$ character of the
$dx^a dx^b$ error terms follows from the totally geodesic
character of $\ml$. We shall need an anti-symmetry property of the
derivatives of the $g_{aA}$'s, which we now derive: by
construction, the coordinate rays $s\to sx^A$ are affinely
parameterised geodesics. This gives
$$0 = \frac {d^2x^a}{ds^2} + \Gamma^a_{ij} \frac {dx^
i}{ds}\frac {dx^j}{ds} =\Gamma^a_{BC} \frac {dx^A}{ds}\frac
{dx^B}{ds}\;.$$ Since the vector $dx^A/ds$ can be arbitrarily
chosen at $s=0$ this implies \bel{afp0a}0=\Gamma^a_{BC}|_{x^A=0}
\quad \Longleftrightarrow \quad (g_{aA,B} +
g_{aB,A})|_{\{x^C=0\}}=0\;,\ee where a comma denotes a partial
derivative. (Similar arguments may of course be  used to justify
the $O(\rho^2)$ character of the remaining error terms in
\eq{afp0}.)

Exponentiating \eq{me4} shows that on each space $(T_q\ml)^\perp$
the one-parameter group of diffeomorphisms $\phi_s$ generated by
 $X$ acts as a rotation of angle $\kappa_i s$ of
the planes $\mathrm{Vect}\{\partial_{2i-1},\partial_{2i}\}$, $1\le
i\le \ell$. The definition of geodesic coordinates implies that on
$\mcU_p$ the Killing vector $X$ equals \bel{afp1}
X=\sum_{i=1}^{\ell}\kappa_i (x^{2i-1}\partial_{2i}-
x^{2i}\partial_{2i-1})\;.\ee  This equation is exact; there are no
error terms, as opposed to \emph{e.g.}\/ \eq{afp0}. Since
$\phi_{2\pi}$ is the identity we have $\kappa_i\in\Z^*$, and since
almost all orbits have period $2\pi$ it follows that at least one
$|\kappa_i|$
--- say $|\kappa_1|$
--- equals one.
For $\ell\ge 2 $ by renaming and multiplication by $-1$ of the
coordinates one can arrange to have
\bel{me6} 1=|\kappa_1|\le \kappa_i\le \kappa_{i+1}\le
\kappa_\ell\;,\quad 2\le i \le \ell-1\;,\ee and we will always
assume that \eq{me6} holds. We have assumed that an orientation of
$(T_q\ml)^\perp$ has been chosen, and the sign in $\kappa_1$ is
chosen so that the coordinates of \eq{afp1} have the correct
orientation.
Continuity shows that 
the $\kappa_i$'s are constant over each connected component of
$\ml$.

We shall denote by $\rho_i$ and $\varphi_i$  the polar coordinates
in the $(x^{2i-1},x^{2i})$ planes: \bel{afp3}
x^{2i-1}=\rho_i\cos\varphi_i\;,\qquad
x^{2i}=\rho_i\sin\varphi_i\;,\ee so that\bel{afp5}
X=\sum_{i=1}^{\ell}\kappa_i \frac
{\partial}{\partial{\varphi_i}}\;.\ee If all the $\kappa_i$'s are
ones, then all orbits in $\mcU_p$ have period $2\pi$, in which
case
$$\miso\cap\mcU_p=\pi_\Sigma^{-1}(\sings)\cap\mcU_p=\emptyset\;.$$
Otherwise $\ell\ge 2$ and there exists a smallest $i_2$ such that
$\kappa_i\ge 2$ for $i\ge i_2$. If $q$ is such that $\rho_i(q)=0$
for $1\le i< j$, and $\rho_{j}(q)>0$, then the orbit of $X$
through $p$ has period $2\pi/\kappa_j$. It follows that  an  orbit
through $q\in\mcU_p$ has trivial isotropy if and only if
$$\sum_{i=1}^{i_2-1}\rho_i(q)\ne 0 \;.$$  We have shown:

\begin{Proposition}\label{PB1}We have
 $$\ml\cap \overline{\miso} \ne \emptyset \quad \Longleftrightarrow \quad \exists\; i\ \mbox{\rm such that}\ \kappa_i\ge 2\;,$$
in particular
$$\mtwo\cap \overline{\miso}=\emptyset \;.$$
\end{Proposition}

 To proceed further, we need to understand
 the structure of $\Sigma$ near $\pi_\Sigma\ml$. We first use the
  polar coordinates \eq{afp3}, and then introduce new
angular variables $\varphi,\psi_i$, parameterising
$\underbrace{S^1\times\cdots\times S^1}_{n \ \mathrm{factors}}$,
  defined as \bel{af14} \varphi:= \varphi_1\;,\qquad
\psi_i := \varphi_i- \kappa_1\kappa_i \varphi_1\;, \
i=2,\ldots,n\;,\ee so that, using \eq{afp5},
$$ X(\varphi)=1\;,\qquad X(\psi_i)=X(\rho_i)=0\;.$$
It follows that $X=\partial_\varphi$, and that
$(\rho_1,(\rho_i,\psi_i)_{i\ge 2})$,  can be used as local
coordinates on $\zSigma$.
 There is a usual ``polar coordinates
singularity" at  the sets $\{\rho_i=0, u\ne 0\}$ for $i\ge 2$. As
already pointed out, for $i$'s such that $\kappa_{i+1}>\kappa_i$
the periodicity of the $\varphi_i$'s jumps down from
$2\pi/\kappa_i$ to $2\pi/\kappa_{i+1}$ at the sets
$\{\rho_1=\ldots=\rho_{i}=0,\; u\ne 0\}$. This leads to an
identical jump of the periodicity of the $\psi_i$'s, leading to
orbifold singularities of increasing complexity at each of those
sets. In conclusion, within the domain of the coordinate system
$(\rho_1,(\rho_i,\psi_i)_{i\ge 2})$ the differentiable part
$\zSigma$ of $\Sigma$ takes the form
$$\{u>0\} $$ if all the $\kappa_i$'s are ones, and
$$\{u>0\}\setminus\{\rho_{1}=\ldots=\rho_{i_2}=0\} $$
otherwise. Further, $(\rho_1,(\rho_i,\psi_i)_{i\ge 2})$ provide a
well behaved coordinate system of polar type  on $\zSigma$ in a
neighborhood of $\pi_\Sigma\ml$.

 \Eqsone{afp0} and \eq{afp1} imply
  \bel{afp7}\xi:=g(X,\cdot)=
\sum_{i=1}^{\ell}\kappa_i (x^{2i-1}dx^{2i}- x^{2i}dx^{2i-1}) +
\mnu_adx^a + \sum_{i}O(\rho^3)dx^i \;,\ee where \bel{afp7a}
\mnu_a:=  g_{aA,B}|_{\{x^C=0\}}X^Ax^B\;.\ee  At this stage it is
adequate to
 enquire about the geometric character of the objects
defined so far. Note that the locally defined coordinates $x^A$
appearing in \eq{afp0} are only determined modulo $x^a$--dependent
rotations: \bel{afp7a0} x^A \to \bar x^A:=\omega^A{}_B(x^a)
x^B\;,\ee where, at each $x^a$, $\omega^A{}_B$ is an $2\ell$ by
$2\ell$ orthogonal matrix that preserves all the spaces
$\mathrm{Vect}\{\partial_{2i-1},\partial_{2i}\}$. Suppose, thus,
that two coordinate systems $(\bar x^A,\bar x^a)$ and $(x^A,x^a)$
are given, with $\bar x^a=x^a$, and with $\bar x^A$ related to
$x^A$ via \eq{afp7a0}. It is convenient to put bars on $g_{AB}$,
$\nu_a$, \emph{etc},\/ to denote those objects in the barred
coordinate system. One easily finds the following transformation
law under \eq{afp7a0}:\bel{afp7a1} \frac{\partial \bar
g_{aA}}{\partial \bar x^B}|_{\{\bar x^C=0\}} \to \frac{\partial
g_{aA}}{\partial x^B}|_{\{ x^C=0\}}=\sum_D\omega^D{}_A\left(
\frac{\partial \bar g_{aD}}{\partial \bar x^E}|_{\{\bar x^C=0\}}
\omega^E{}_B+ \omega^D{}_{B,a}\right)\;.\ee It follows that
\bel{afp7a3} \bar\mnu_a \to \mnu_a=\bar\mnu_a +\sum_D
\omega^D{}_{B,a}\omega^D{}_A X^Ax^B\;.\ee Now, $\omega$ is a
block-diagonal matrix consisting of two-by-two blocks, each of
them of the form $$ \left[\begin{array}{cc} \cos (\theta_i(x^a)) &
-\sin (\theta_i(x^a))\\ \sin (\theta_i(x^a)) & \cos
(\theta_i(x^a))
\end{array}\right]\;.$$ Inserting this into \eq{afp7a3} one obtains
\bel{afp7a4} \bar\mnu_a \to \mnu_a=\bar\mnu_a +\sum_{i=1}^\ell
\kappa_i \left((x^{2i-1})^2+(x^{2i})^2\right)
\frac{\partial\theta_i}{\partial x^a}\;.\ee So far we have assumed
that $\bar x^a= x^a$; this last restriction is removed in a
straightforward way, leading to a tensorial transformation law of
the right-hand-side of \eq{afp7a4} under the transformation
$$(\bar x^A,\bar x^a)\to (\tilde x^A=\bar x^A,\tilde
x^a=\phi^a(\bar x^b))\;.$$ It should be emphasised that in general
we will not be able to achieve $\omega= \mathrm{id}$ when going
from one coordinate patch $x^a$ to another on $\ml$. This implies
that $\mnu_adx^a$ does \emph{not} transform as a one-form when
passing from one $x^a$--coordinates patch on $\ml$ to another,
except in the case in which the $x^A$'s can be globally
``synchronised" over $\ml$
---  this occurs if and only if each of the bundles $\mathrm{Vect}\{\partial_{2i-1},\partial_{2i}\}$ is
trivial.

In order to evaluate the $\psings(\rho)$ integral in \eq{nke18a}
we need to calculate $u^{-3}*_{g_\Sigma}\omega=u^{-3}\lambda$,
with $\lambda$ being defined as the $\Sigma$--equivalent of the
two--form $\hlambda$ of \eq{nke0}. The simplest way of doing this
proceeds via the calculation of the form $\beta$ of \eq{nke0.4},
\emph{cf.}\/ \eq{nke0.6}.
 Consider,
first, the case $\ell=1$, set
\bel{afp4a} \gamma_a:=g_{a1,2}|_{\{x^C=0\}}\;;\ee the
anti-symmetry property \eq{afp0a} gives
$$\mnu_a=\gamma_a \rho_1^2\;,$$
hence
 \bel{afp7b}\xi=\kappa_1
\rho_1^2d\varphi +  \gamma_a\rho_1^2dx^a + \sum_{i}O(\rho^3)dx^i
\;,\ee so that \bel{af[7c} \beta:=u^{-2}\xi = \kappa_1d\varphi +
\gamma_a dx^a + O(\rho)d\rho_1
+\sum_{a}O(\rho)dx^a\;.\ee \Eq{afp7a4} shows that $\gamma_adx^a$
is a connection form on the $U(1)$--principal bundle of unit
vectors normal to $\mtwo$: \bel{afp7n4} \bar\gamma_a \to
\gamma_a=\bar\gamma_a + \frac{\partial\theta_1}{\partial
x^a}\;.\ee In particular the curvature two-form $$F=d\gamma$$ is a
well-defined two-form on $\mtwo$.

Let us return to \eq{afp7}-\eq{afp7a} for general $\ell\ge 2$;
\Eq{af14} gives
  \bean\xi& = &  \kappa_1\rho_1^2d\varphi+
\sum_{i=2}^{\ell}\kappa_i
\rho_i^2(d\psi_i+\kappa_1\kappa_id\varphi) + \mnu_adx^a +
\sum_{i}O(\rho^3)dx^i
\\ &=& u^2\Big(\kappa_1d\varphi +
\sum_{i=2}^{\ell}\kappa_i u^{-2}\rho_i^2d\psi_i+ u^{-2}\mnu_a dx^a
\nonumber
\\ && 
+ \sum_{i\ge 1}O(\rho)d\rho_i+\sum_{i\ge
2}O(\rho^2)d\psi_i+\sum_{a}O(\rho)dx^a\Big)\;,\eeal{afp7b1} with
 \bea
u&=&\sqrt{\sum_{i=1}^\ell
\kappa_i^2\rho_i^2}+O(\rho^3)\;.\eeal{afp4a.1}
\Eqs{nke0.4}{nke0.6} together with \eq{nke8} and \eq{afp7b1}
 immediately lead to \bean u^{-3}*_{g_\Sigma}\omega & = & u^{-3}\lambda
 \\ &= &
 \sum_{i=2}^\ell
d\left(\frac{\kappa_i\rho_i^2}{u^2}d\psi_i \right) +
d(u^{-2}\mnu_a dx^a)
\nonumber
\\ && 
+ \sum_{i,j\ge 1}O(1) d\rho_j\wedge d\rho_i+\sum_{i\ge 2,j\ge
1}O(\rho)d\rho^j\wedge d\psi_i+\sum_{i,j\ge 2\ge
1}O(\rho^2)d\psi_j\wedge d\psi_i\nonumber\\
&&+\sum_{a,i}O(1)d\rho_i\wedge dx^a
+\sum_{a,j}O(\rho)d\psi_j\wedge dx^a+\sum_{a,b}O(\rho)dx^a\wedge
dx^b\;, \nonumber \\ &&\eeal{afp17} which is used in the proof of
Theorem~\ref{T4}.

We note  the following necessary condition for staticity:
\begin{Proposition}\label{PA1} If $(M,g)$ is static, then
$\ml=\emptyset$ for $\ell> 1$.
\end{Proposition}

\proof Calculating directly from \eq{afp7} we find \beaa d\xi&=&
-2\sum_{i=1}^{\ell}\kappa_i \,dx^{2i-1}\wedge
dx^{2i}+d\mnu_a\wedge dx^a+ \sum_{i}O(\rho^2)dx^i\;,\eeaa
so that
 \bean d\xi\wedge \xi &=&
-2\sum_{i\ne j=1}^{\ell}\kappa_i\kappa_j \,dx^{2j-1}\wedge
dx^{2j}\wedge(x^{2i-1}dx^{2i}- x^{2i}dx^{2i-1}) \nonumber \\ && +
\sum_{A,B,a}O(\rho^2)dx^A\wedge dx^B\wedge dx^a+
\sum_{i,j,k}O(\rho^3)dx^i\wedge dx^j \wedge dx^k\;,\nonumber \\ &&
\eeal{afp29} which clearly does never vanish  when $\ell\ge 2$ on
a sufficiently small neighborhood of $\ml$. \qed

\section{A family of non-degenerate black hole
solutions}\label{Sndbhs}
\subsection{An injectivity theorem}

We start by proving point (a) of Theorem~\ref{thm:einstein2}:

\begin{theor}\label{kmin}
Let $K_{\max} (x)$ and $K_{\min}(x)$ denote the largest and the
smallest sectional curvature of $g$ at $x$. If for all $x\in M$ it
holds that either $K_{max}(x)\leq 0$ or $K_{min}(x)\geq
-2(n-1)/n$, then  the operator $\Delta_L+2(n-1)$ has trivial $L^2$
kernel.
\end{theor}

\proof  We use the notations of Lee~\cite{Lee:fredholm}, except
that we work in dimension $n$, not $n+1$. For all $x\in M$, let
$$
a(x)=\sup\{(\mathring{Rm}_xh_x,h_x)/|h_x|^2\;,\;h\in S^2_0\}.
$$
From~\cite[Lemma~12.71]{Besse}  we have that
$$
a(x)\leq \min \{(n-2)K_{max}(x)+n-1,-(n-1)-nK_{min}(x)\}\;,
$$
showing that under the current hypotheses we have $n-1-a(x)\ge 0$.
 The proof
in~\cite[p.~67]{Lee:fredholm} establishes then that the operator
$\Delta_L+2(n-1)$ has trivial kernel. \qed

\subsection{Sectional curvatures of generalised Kottler metrics}
 We consider a generalised Kottler metric,
\bel{goodm} g=\frac{1}{V(r)}dr^2+V(r)d\theta^2+r^2\hat{g}\;,\ee
where $\hat{g}=\hat{g}_{AB}dx^Adx^B$ does not depend on $r$ and
$\theta$. The non-trivial components of the Riemann tensor are
$$
R_{r\theta r\theta}=-\frac{1}{2}V''$$
$$
R_{rArB}=-\frac{rV'}{2V}\hat{g}_{AB},
$$
$$
R_{\theta A \theta B}=-\frac{r^{}V'V}{2}\hat{g}_{AB},
$$
$$
R_{ABCD}=r^2\hat{R}_{ABCD}-r^2V(\hat{g}_{AC}\hat{g}_{BD}-\hat{g}_{AD}\hat{g}_{BC}).$$
In particular if  $V(r)=c +r^2-2mr^{-(n-3)}$, we obtain
$$
R_{r\theta r\theta}=-1+(n-3)(n-2)mr^{-(n-1)},$$
$$
R_{rArB}=-[(r^2+m(n-3)r^{-(n-3)})/V]\hat{g}_{AB},
$$
$$
R_{\theta A \theta B}=-(r^2+m(n-3)r^{-(n-3)})V\hat{g}_{AB},
$$
$$
R_{ABCD}=r^2\hat{R}_{ABCD}-r^2V(\hat{g}_{AC}\hat{g}_{BD}-\hat{g}_{AD}\hat{g}_{BC}).$$
Let $U=(U^r,U^\theta,U^A)=(U^r,U^\theta,\hat{U})$ and
$W=(W^r,W^\theta,W^A)=(W^r,W^\theta,\hat{W})$ be two orthogonal
vectors with norm $1$, the sectional curvature of span$(U,W)$ is
$$
\begin{array}{lll}
K(U,W)&=&\displaystyle{-\frac{1}{2}V''(r)[U^rW^\theta-W^rU^\theta]^2}\\
&&\displaystyle{-\frac{rV'(r)}{2}\left\{\frac{1}{V(r)}\|U^r\hat{W}-W^r\hat{U}\|^2_{\hat{g}}
+{V(r)}\|U^\theta\hat{W}-W^\theta\hat{U}\|^2_{\hat{g}}\frac{}{}\right\}}\\
&&+\displaystyle{r^2\hat{R}_{ABCD}U^AW^BU^CW^D-r^2V(r)(\|\hat{U}\|^2_{\hat{g}}\|\hat{W}\|^2_{\hat{g}}
-\langle \hat{U},\hat{W}\rangle^2_{\hat{g}})}\\
&=&\displaystyle{-\frac{1}{2}V''(r)[U^rW^\theta-W^rU^\theta]^2}\\
&&\displaystyle{-\frac{rV'(r)}{2}\left\{\frac{1}{V(r)}\|U^r\hat{W}-W^r\hat{U}\|^2_{\hat{g}}
+{V(r)}\|U^\theta\hat{W}-W^\theta\hat{U}\|^2_{\hat{g}}\frac{}{}\right\}}\\
&&+\displaystyle{r^2(\hat{K}(\hat{U},\hat{W})-V(r))(\|\hat{U}\|^2_{\hat{g}}\|\hat{W}\|^2_{\hat{g}}
-\langle \hat{U},\hat{W}\rangle^2_{\hat{g}})}\;.
\end{array}
$$
Set
$$(a,b,c)=(V^{-1/2}(r)U^r,V^{1/2}(r)U^\theta,r\hat{U})\;,$$
and
$$(x,y,z)=(V^{-1/2}(r)W^r,V^{1/2}(r)W^\theta,r\hat{W})\;,$$
so that $a^2+b^2+|c|^2=x^2+y^2+|z|^2=1$ and $ax+by+\langle
c,z\rangle=0$, where the norm $|\cdot |$ and the scalar product
$\langle\cdot,\cdot\rangle $ are taken with respect to $\hat{g}$.
We can rewrite the sectional curvature as
$$
\begin{array}{lll}
K(U,W)
&=&\displaystyle{-\frac{1}{2}V''(r)[ay-bx]^2}\\
&&\displaystyle{-\frac{r^{-1}V'(r)}{2}\left\{|az-xc|^2+|bz-yc|^2\right\}}\\
&&+\displaystyle{r^{-2}[\hat{K}(\hat{U},\hat{W})-V(r)](|c|^2|z|^2-\langle
c,z\rangle^2)}\;.
\end{array}
$$
Setting
$$k=\min(-\frac{1}{2}V''(r),-\frac{r^{-1}V'(r)}{2},r^{-2}[\hat{K}(\hat{U},\hat{W})-V(r)])\;,$$
we obtain
$$ K(U,W)\geq
k\left([ay-bx]^2+|az-xc|^2+|bz-yc|^2+|c|^2|z|^2-\langle
c,z\rangle^2\right)=k\;.
$$
Similarly, $$ K(U,V)\le K:=
\max(-\frac{1}{2}V''(r),-\frac{r^{-1}V'(r)}{2},r^{-2}[\hat{K}(\hat{U},\hat{W})-V(r)])\;.$$
Coming back to $V(r)=c +r^2-2mr^{-(n-3)}$, we further assume that
$\hat{K}(\hat{U},\hat{W})=c$ and $n\geq4$. One then finds
$$k = -1 + r^{-(n-1)}\min \left\{(n-3)(n-2)m, - (n-3)m,
2m\right\}\;,$$ so that
 if $m\geq0$ then
\bel{mpos} k=-1-m(n-3)r^{-(n-1)}\;,\quad
K=-1+m(n-3)(n-2)r^{-(n-1)}\;. \ee while for $m\leq0$ one has
\bel{mneg}
k=-1+m(n-3)(n-2)r^{-(n-1)}=-1-|m|(n-3)(n-2)r^{-(n-1)}\;, \ee
\bel{mneg2}K=-1+|m|(n-3)r^{-(n-1)}\;. \ee {\bf A)} If $m\geq 0$,
we have $k\geq -2(n-1)/n$ if and only if \bel{mposcond}
mr_+^{-(n-1)}\leq\frac{n-2}{n(n-3)}\;,\ee where $r_+$ is the
unique positive solution of \bel{rooteq}V(r_+)=0\quad
\Longleftrightarrow \quad mr_+^{-(n-3)} = \frac 12\left(c
+r_+^2\right)\;.\ee On the other hand, $K\le 0$ will hold if and
only if \bel{mposcond2} mr_+^{-(n-1)}\leq\frac1{(n-3)(n-2)}\;. \ee
Since the right-hand-side of \eq{mposcond} is larger than that of
\eq{mposcond2} for $n>4$, with equality for $n=4$, the former
condition is less restrictive than the latter. For further use we
note that \bel{mcal}mr_+^{-(n-1)}= r_+^{-2}mr_+^{-(n-3)}= \frac 12
\left( 1+\frac c {r_+^2}\right)\;.\ee

{\bf a)} For $c=1$  the left-hand-side of \eq{mposcond} is
strictly larger than $1/2$ for $m>0$ by \eq{mcal}, while the
right-hand-side is less than or equal to $1/2$ when $n\ge 4$, and
our non-degeneracy criterion in terms of $k$ does not apply.
Similarly one finds that some sectional curvatures are always
positive at $r=r_+$.

If we assume that the sectional curvatures of $\hat g$ are equal
to $c=1$, and that the manifold $N^{n-2}$ carrying the metric
$\hat g$ is compact, then $(N^{n-2},\hat g)$ is clearly of
positive Yamabe type, and one can likewise attempt to use point
(b) of Theorem~\ref{thm:einstein2} to prove non-degeneracy.
Unfortunately, it turns out that  the sectional curvature
inequality there is always violated at $r_+$.

 {\bf b)} If $c=0$ then $r_+= (2m)^{1/(n-1)}$, giving $1/2=1/2$ for $n=4$ in \eq{mposcond}, without restrictions on
 $m$. However, for $n\ge 5$ the right-hand-side of \eq{mposcond} is always
 smaller than one half. 
Similarly the inequality of point (b) of
Theorem~\ref{thm:einstein2} always fails.

{\bf c)} If $c=-1$ then the map $$[1,\infty)\ni
r_+(m)\longleftrightarrow\; m(r_+)\in [0,\infty)$$ is a bijection,
and for all $ 0\le m\le m_+ $ from \eq{mposcond} we obtain
non-degeneracy, where $m_+=\infty$ if $n=4$. For $n\ge 5$ the
value of $m_+$ can be found by first solving \eq{mposcond} in
terms of $r_+$ using \eq{mcal},
$$ r_+(m_+)=
\sqrt{{{\frac{n(n-3)}{n(n-3)-2(n-2)}}}}=\sqrt{{{\frac{n(n-3)}{(n-1)(n-4)}}}}\;.$$
\Eq{rooteq} can then be used to calculate $m_+=m_+(n)$:
\bel{mplus}
m_+(n)=\left\{%
\begin{array}{ll}
    \infty, & \hbox{$n=4$;} \\
\frac{(n-2)}{(n-1)(n-4)}\left(\frac{n(n-3)}{(n-1)(n-4)}\right)^{\frac{n-3}{2}}, & \hbox{$n\ge 5$.} \\
\end{array}%
\right.\ee

{\bf B)} If $m< 0$, we have $k\geq -2(n-1)/n$ if and only if
\bel{kcond2} |m|r_+^{-(n-1)}\leq\frac{1}{n(n-3)}\;, \ee while
 $K\leq 0$ is equivalent to \bel{Kcond2}
|m|r_+^{-(n-1)}\leq\frac{1}{(n-3)}\;,\ee this last condition being
less restrictive than \eq{kcond2}.

The only case of interest is $c=-1$, as $V$ has no zeros
otherwise. The map
$$\left[r_{\min}:=\sqrt{\frac{n-3}{n-1}},1\right]\ni r_+(m)\longleftrightarrow m(r_+)\in
\left[\frac 12 (r^{n-1}_{\min}-r^{n-3}_{\min}),0\right]$$ is a
bijection, and for all $ m_-\le m<0 $ from \eq{Kcond2} we obtain
negative sectional curvatures, where \bel{rplusmi}
r_+(m_-)=r_{\min}= \sqrt{{{\frac{n-3}{n-1}}}}\;,\ee
\bel{mminus}m_-=m_-(n)=-\frac{1}{n-1}\left(\frac{n-3}{n-1}\right)^{\frac{n-3}{2}}\;.\ee
Recall that $r_{\min}$ given by \eq{rplusmi} corresponds to the
smallest value of $r_+(m)$ for which a regular solution exists.
Equations~\eq{rplusmi}-\eq{mminus} show that non-degeneracy holds
in the whole range of negative masses compatible with a
singularity-free metric. Summarising, we have proved:
\begin{Proposition} Let $V(r)=-1 +r^2-2mr^{-(n-3)}$, suppose
that $\hat g$ is a metric of constant sectional curvature equal to
$-1$ on a compact manifold $N^{n-2}$, then for $n\ge 4$ and
for\footnote{The case $m=m_-(n)$ corresponds to $V'(r_{\min})=0$,
which leads to a cylindrical end for the metric \eq{goodm}, and is
therefore excluded by the requirement that the associated
Riemannian manifold be conformally compact. The corresponding
Lorentzian solution is a regular black hole with vanishing surface
gravity.} $m\in (m_-(n),m_+(n)]$, as given by \eq{mminus} and
\eq{mplus}, the metric \eq{goodm} is non degenerate. In dimension
four all singularity-free such solutions are non-degenerate.
\end{Proposition}



\bibliographystyle{amsplain}
\bibliography{
../references/newbiblio,%
../references/reffile,%
../references/bibl,%
../references/Energy,%
../references/hip_bib,%
../references/netbiblio,%
../references/newbib}
\end{document}